\documentclass[letterpaper]{article} 
\usepackage{aaai2026}  
\usepackage{times}  
\usepackage{helvet}  
\usepackage{courier}  
\usepackage[hyphens]{url}  
\usepackage{graphicx} 
\usepackage{natbib}  
\usepackage{caption} 
\usepackage{algorithm}
\usepackage{algorithmic}

\usepackage{booktabs}       
\usepackage{multirow}
\usepackage{multicol}
\usepackage{colortbl}

\usepackage{utfsym}

\usepackage{booktabs}
\usepackage{multirow}
\usepackage{graphicx}
\usepackage{enumitem}

\usepackage{amsmath} 
\usepackage{newfloat}
\usepackage{listings}
\DeclareCaptionStyle{ruled}{labelfont=normalfont,labelsep=colon,strut=off} 
\floatstyle{ruled}
\newfloat{listing}{tb}{lst}{}
\floatname{listing}{Listing}
\title{TFRank: Think-Free Reasoning Enables Practical Pointwise LLM Ranking}
\author{
Yongqi Fan\textsuperscript{\rm $\diamondsuit \spadesuit$\thanks{~~Co-first authors, work was done when interning at Tencent.}},
 Xiaoyang Chen\textsuperscript{\rm $\heartsuit \clubsuit \spadesuit$\footnotemark[1]},
 Dezhi Ye\textsuperscript{\rm $\spadesuit$},
 Jie Liu\textsuperscript{\rm $\spadesuit$},
 Haijin Liang\textsuperscript{\rm $\spadesuit$} \\
 Jin Ma\textsuperscript{\rm $\spadesuit$},
 Ben He\textsuperscript{\rm $\heartsuit \clubsuit$},
 \textbf{Yingfei Sun}\textsuperscript{\rm $\heartsuit$},
 \textbf{Tong Ruan}\textsuperscript{\rm $\diamondsuit$}\thanks{~~Corresponding author.} \\
}
\affiliations{
\textsuperscript{\rm $\diamondsuit$}East China University of Science and Technology, Shanghai, China \\
\textsuperscript{\rm $\spadesuit$}Tencent, \textsuperscript{\rm $\heartsuit$}University of Chinese Academy of Sciences \\
\textsuperscript{\rm $\clubsuit$}Chinese Information Processing Laboratory, Institute of Software, Chinese Academy of Sciences \\
\texttt{johnnyfans@mail.ecust.edu.cn}, \texttt{chenxiaoyang19@mails.ucas.ac.cn} \\
\texttt{dezhiye@tencent.com}, \texttt{\{benhe, yfsun\}@ucas.ac.cn}, \texttt{ruantong@ecust.edu.cn} \\
}

\usepackage{bibentry}

\begin{document}

\maketitle

\begin{abstract}
Reasoning-intensive ranking models built on Large Language Models (LLMs) have made notable progress. However, existing approaches often rely on large-scale LLMs and explicit Chain-of-Thought (CoT) reasoning, resulting in high computational cost and latency that limit real-world use. 
To address this, we propose \textbf{TFRank}, an efficient pointwise reasoning ranker based on small-scale LLMs. 
To improve ranking performance, TFRank effectively integrates CoT data, fine-grained score supervision, and multi-task training. Furthermore, it achieves an efficient ``\textbf{T}hink-\textbf{F}ree" reasoning capability by employing a ``think-mode switch'' and pointwise format constraints. Specifically, this allows the model to leverage explicit reasoning during training while delivering precise relevance scores for complex queries at inference without generating any reasoning chains.
Experiments show that TFRank achieves performance comparable to models with four times more parameters on the BRIGHT benchmark, and demonstrates strong competitiveness on the BEIR benchmark. Further analysis shows that TFRank achieves an effective balance between performance and efficiency, providing a practical solution for integrating advanced reasoning into real-world systems.
\end{abstract}

\section{Introduction}\label{sec:intro}

Driven by the advancements in Large Language Models (LLMs), Information Retrieval (IR) systems are increasingly confronted with the demand to handle complex, inferential user queries~\cite{deepseek-r1,qwen3,bright,followir}. In modern applications such as Retrieval-Augmented Generation (RAG) systems and Multi-Agent frameworks~\cite{rankrag,sos,multiagents}, the ability to perform reasoning over both queries and retrieved documents is essential for ranking models to deliver accurate and relevant information. This has led to research on \textit{ reasoning-intensive ranking}, which aims to develop ranking algorithms capable of understanding subtle relationships and accurately assessing the relevance of documents to complex information needs.
\begin{figure*}[ht]
\centering
\includegraphics[width=0.83\linewidth]{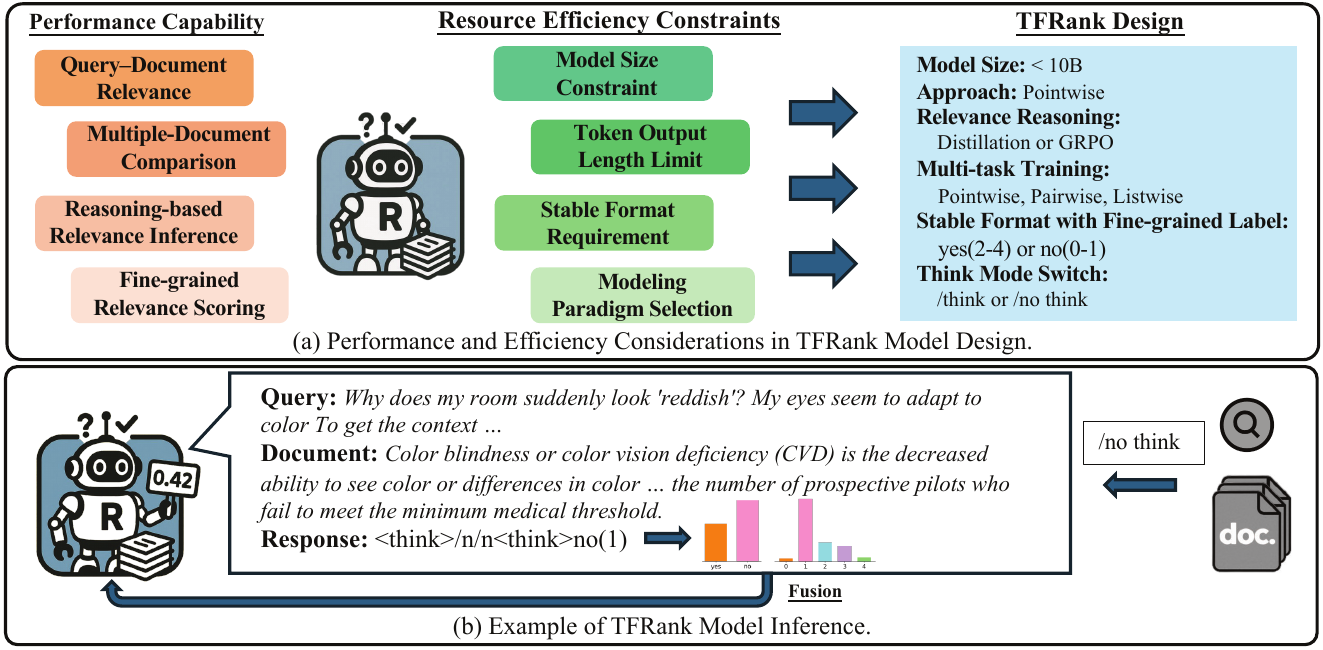} 
\caption{
Overview of the TFRank framework. (a) summarizes key performance and efficiency considerations in the model design. (b) provides an example of inference with the TFRank model.}
\label{fig:framework}
\end{figure*}

Early attempts to endow rankers with reasoning capabilities relied primarily on the power of LLMs. Methods leverage LLMs’ instruction‐following abilities to flexibly judge query–document relevance~\cite{rankgpt,lpl,rankrag}. To further enhance performance on complex queries, many approaches explicitly incorporate the Chain-of-Thought (CoT) mechanism. By generating reasoning text before making a final relevance judgment, models such as Rank1 \cite{weller2025rank1}, Rank-R1 \cite{zhuang2025rank}, and REARANK \cite{zhang2025rearank} have demonstrated significant performance improvements in handling queries that require sophisticated reasoning.

However, the pursuit of performance often incurs prohibitive costs, hindering the transition from research prototypes to deployable systems.   
A number of state-of-the-art rankers rely on computationally and financially expensive models, whether proprietary (e.g., GPT-4) or massive open-source versions (e.g., Llama-3-70B-Instruct~\cite{llama3}). Moreover, the reasoning process that boosts performance is token-intensive, leading to significant inference latency. Finally, listwise or setwise approaches, adopted for cross-document comparison, exhibit limited scalability, rendering them unsuitable for low-latency, real-time ranking.

Therefore, as illustrated in Figure~\ref{fig:framework}, the core challenge in designing a practical, reasoning-capable ranker is: \textbf{How can we leverage the advantages of reasoning without sacrificing efficiency and robustness in real-world scenarios?} A deployable ranker must meet two essential requirements. It should offer strong performance, including robust query-document relevance modeling, multi-document comparison, reasoning-based inference, and fine-grained scoring. It must also satisfy strict efficiency constraints, such as compact model size, short output length, stable formatting, and a scalable pointwise ranking architecture.

To address this challenge, we propose a novel training and inference pipeline for \emph{pointwise} ranking that balances efficiency and performance with small-scale LLMs. 
In the training phase, we adopt a multi-task strategy that integrates pointwise, pairwise, and listwise supervision with fine-grained relevance labels and CoT signals, enabling the model to learn nuanced and precise scoring. 
To further strengthen the model's capability, we incorporate an optional GRPO~\cite{shao2024deepseekmath} training for advanced reasoning optimization. 
At inference, a ``think-mode switch'' and pointwise output constraints prompt the model to generate direct relevance scores for each query--document pair, fully bypassing explicit reasoning generation.

Experiments on the reasoning-intensive BRIGHT benchmark reveal a striking finding: our ``Think-Free'' inference mode, which omits explicit reasoning chains, not only dramatically improves efficiency but also surpasses the ranking performance of the full-reasoning mode. We term this phenomenon \textbf{``Think-Free" Reasoning} and name our model \textbf{TFRank} accordingly. 
This suggests the model internalizes its reasoning abilities via multi-task training, enabling it to produce high-quality judgments directly. The efficacy of TFRank is particularly pronounced on smaller models: for instance, our 1.7B TFRank model competes with 7B-parameter baselines on BRIGHT, models over four times its size. Moreover, on the general BEIR benchmark, TFRank maintains comparable performance to existing methods while offering substantial efficiency gains. Thus, TFRank provides a validated and efficient pathway for building a new generation of ranking systems that combine state-of-the-art reasoning capabilities with production-level feasibility.

Our contributions can be summarized as follows:
\begin{itemize}
    \item We propose \textbf{TFRank}\footnote{We have open-sourced our code and data at: \url{https://github.com/JOHNNY-fans/TFRank}}, a novel and efficient framework that enables small-scale LLMs to perform reasoning-intensive ranking by internalizing reasoning patterns through multi-task training.
    \item We identify the \textbf{``Think-Free'' Reasoning} phenomenon, where suppressing explicit reasoning chains at inference leads to both superior ranking accuracy and dramatic efficiency gains.
    \item Our experiments demonstrate that TFRank achieves promising results on the BRIGHT benchmark, with a 1.7B model competing with 7B baselines, proving its practical viability for production systems.
\end{itemize}
\section{Related Works}
\paragraph{Application of LLMs in Ranking}
LLM-based ranking methods can be categorized into three paradigms according to how documents are processed. The \textbf{pointwise} paradigm evaluates each document independently for query relevance and outputs a score~\cite{rankllama, prorank, weller2025rank1, qwenranker, liu2024demorank}, offering high parallelism and low latency, which makes it suitable for real-time applications. The \textbf{pairwise} paradigm compares document pairs to provide finer-grained supervision~\cite{pairwise1, pairwise2}, but its quadratic complexity limits scalability with many candidates. \textbf{Listwise} and \textbf{setwise} methods~\cite{rankgpt, lpl, setwise} process multiple documents jointly, but face increased input/output length and lower efficiency due to sliding-window strategies, hindering real-time deployment. Recent studies further explore long-context LLMs to fully rank candidate passages~\cite{liu-etal-2025-sliding}, which improves efficiency over sliding windows but inevitably increases latency. In contrast, our proposed TFRank adopts an efficient pointwise architecture while integrating strong reasoning abilities through targeted training.

\begin{table*}[ht]
    \centering \small
    \setlength{\tabcolsep}{1mm}
    \begin{tabular}{lcc|ccccccc|cc|ccc|c}
        \toprule
        \multicolumn{2}{c}{\multirow{2}{*}{\centering\textbf{Model}}} 
        & \multirow{2}{*}{\centering\textbf{Approach}}
        & \multicolumn{7}{c|}{\textbf{StackExchange}} 
        & \multicolumn{2}{c|}{\textbf{Coding}} 
        & \multicolumn{3}{c|}{\textbf{Theorem-based}} 
        & \multirow{2}{*}{\centering\textbf{Avg.}} \\
        \cmidrule(lr){4-10} \cmidrule(lr){11-12} \cmidrule(lr){13-15}
        & & & Bio. & Earth. & Econ. & Psy. & Rob. & Stack. & Sus. 
          & Leet. & Pony 
          & AoPS & TheoQ. & TheoT. 
          & \\
        \midrule
        BM25   & - & Pointwise  & 18.2 & 27.9 & 16.5 & 13.4 & 10.9 & 16.3 & 16.1 & 24.7 & 4.3 & 6.5 & 7.3 & 2.1 & 13.7 \\
        RankGPT-4 & Zero-shot & Listwise & \textbf{33.8} & 34.2 & 16.7 & 27.0 & \underline{22.3} & \textbf{27.7} & 11.1 & 3.4 & 15.6 & 1.2 & 0.2 & 8.6 & 17.0 \\
        Deepseek-R1-rank\_llm\footnotemark & Zero-shot & Listwise & 14.6 & 16.9 & 11.1 & 17.3 & 17.5 & 11.2 & 15.3 & 8.1 & 8.8 & 5.6 & 4.7 & 8.1 & 11.6 \\
        Rank1-7B & SFT & Pointwise & 31.4 & 36.7 & 18.3 & 25.4 & 13.8 & 17.6 & 24.8 & 16.7 & 9.5 & 6.1 & 9.5 & \underline{11.6} & 18.5 \\
        Rank1-14B & SFT & Pointwise & 29.6 & 34.8 & 17.2 & 24.3 & 18.6 & 16.2 & 24.5 & 17.5 & 14.4 & 5.5 & 9.2 & 10.7 & 18.5 \\
        REARANK-7B & GRPO & Listwise & 23.4 & 27.4 & 18.5 & 24.2 & 17.4 & 16.3 & \underline{25.1} & \textbf{27.0} & 8.0 & 7.4 & 7.9 & 9.5 & 17.7 \\
        Rank-R1-3B & GRPO & Setwise & 18.4 & 17.1 & 13.7 & 16.9 & 9.0 & 10.0 & 16.5 & 11.1 & 4.7 & 3.5 & 3.2 & 5.9 & 10.8 \\
        Rank-R1-7B & GRPO & Setwise & 26.0 & 28.5 & 17.2 & 24.2 & 19.1 & 10.4 & 24.2 & 19.8 & 4.3 & 4.3 & 8.3 & 10.9 & 16.4 \\
        Rank-R1-14B & GRPO & Setwise & 31.2 & 38.5 & \underline{21.2} & 26.4 & \textbf{22.6} & 18.9 & \textbf{27.5} & 20.2 & 9.2 & \underline{9.7} & 9.2 & \textbf{11.9} & 20.5 \\
        \midrule
        \multicolumn{15}{c}{\textbf{Llama3.2 Series}}\\
        \midrule
        TFRank-1B & SFT & Pointwise & 23.8 & 25.5 & 8.6 & 15.3 & 7.3 & 11.0 & 8.9 & 9.4 & 9.3 & 4.4 & 6.2 & 2.2 & 11.0 \\
        TFRank-3B & SFT & Pointwise & 31.3 & \underline{42.2} & 19.1 & \underline{27.5} & 14.7 & 17.9 & 19.9 & 19.9 & 2.8 & 6.7 & 9.2 & 8.0 & 18.3 \\
        TFRank-8B & SFT & Pointwise & \underline{31.8} & 41.5 & 19.7 & \textbf{30.0} & 17.2 & 20.9 & 19.2 & \underline{26.0} & \textbf{20.3} & \textbf{10.7} & \underline{10.3} & 9.6 & \textbf{21.4} \\
        \midrule
        \multicolumn{15}{c}{\textbf{Qwen2.5 Series}}\\
        \midrule
        TFRank-0.5B & SFT & Pointwise & 16.7 & 23.0 & 8.3 & 14.4 & 13.7 & 8.1 & 14.1 & 9.0 & 15.2 & 3.4 & 2.8 & 1.1 & 10.8 \\
        TFRank-1.5B & SFT & Pointwise & 22.1 & 23.3 & 15.9 & 20.3 & 11.8 & 13.6 & 14.3 & 17.8 & \underline{17.5} & 6.7 & 9.5 & 6.9 & 15.0 \\
        TFRank-3B & SFT & Pointwise & 29.5 & 36.2 & 16.8 & 24.3 & 15.1 & 13.8 & 18.7 & 23.1 & 8.6 & 8.9 & 9.3 & 9.1 & 17.8 \\
        TFRank-7B & SFT & Pointwise & 31.0 & 41.2 & 20.1 & 26.8 & 19.8 & 18.1 & 22.0 & 23.4 & 16.9 & 7.0 & 10.0 & 10.5 & \underline{20.6} \\
        \midrule
        \multicolumn{15}{c}{\textbf{Qwen3 Series}} \\
        \midrule
        TFRank-0.6B & SFT & Pointwise & 24.8 & 30.0 & 12.0 & 17.5 & 12.9 & 12.1 & 12.7 & 24.4 & 13.1 & 7.1 & \underline{10.3} & 9.8 & 15.6 \\
        TFRank-1.7B & SFT & Pointwise & 25.2 & 29.7 & 17.2 & 26.2 & 15.0 & 16.7 & 17.9 & 21.9 & 10.1 & 4.5 & 7.0 & 9.4 & 16.7 \\
        TFRank-4B & SFT & Pointwise & 31.4 & 40.9 & 19.4 & 26.2 & 18.8 & 19.1 & 20.3 & 23.4 & 13.0 & 7.7 & 10.1 & 9.1 & 20.0 \\
        TFRank-8B & SFT & Pointwise & 29.8 & \textbf{42.3} & \textbf{21.5} & 25.9 & 19.7 & \underline{21.3} & 22.8 & 21.6 & 16.4 & 6.8 & \textbf{10.4} & 9.0 & \underline{20.6} \\
        \bottomrule
    \end{tabular}
    \caption{
        Main evaluation results (NDCG@10) on the BRIGHT benchmark using BM25 as the retriever. 
        TFRank models are trained on their respective backbones. 
        For each column, the highest value is \textbf{bolded} and the second highest is \underline{underlined}.
    }
    \label{tab:bright_yesno_main_pytrec_eval}
\end{table*}
\paragraph{Reasoning-Intensive LLM Ranking}
With the rapid progress of RAG and Multi-Agent systems, as well as the emergence of stronger LLMs (e.g., DeepSeek-R1~\cite{deepseek-r1}, OpenAI o1), explicit reasoning has become increasingly integrated into ranking models~\cite{zhu2023large}. Existing research follows two main directions.
The first is to directly exploit the zero-shot reasoning ability of large LLMs (e.g., Llama-3.1-70B~\cite{llama3}, GPT-4o), as in JudgeRank~\cite{judgerank} and InsertRank~\cite{insertrank}, which have also motivated some more efficient and effective collaborative multi-agent ranking systems~\cite{fan2025llm, liu2025coranking}.
The second direction equips rankers with reasoning through training. This includes CoT distillation methods such as ReasoningRank~\cite{reasoningrank}, Rank1~\cite{weller2025rank1}, and Rank-K~\cite{rankk}, as well as GRPO-based approaches like ReasonRank~\cite{liu2025reasonrank}, REARANK~\cite{zhang2025rearank}, and Rank-R1~\cite{zhuang2025rank}, which optimize reasoning-based ranking on lists or sets~\cite{shao2024deepseekmath}.
However, most of these methods rely on large-scale models and online CoT generation, leading to high inference costs that limit scalability. In contrast, TFRank employs small-scale LLMs and uses CoT only as a training supervision signal. During inference, it bypasses CoT entirely via a ``Think-Free'' mechanism, achieving high efficiency while maintaining performance.
\footnotetext{\url{https://github.com/castorini/rank_llm/tree/main/src/rank_llm}}
\section{Methods}
\subsection{Overall Architecture}
TFRank employs a two-stage architecture. In the \emph{training stage}, we instill reasoning abilities into a small-scale LLM (0.5$\sim$8B) through multi-task supervised fine-tuning (SFT). This process leverages a diverse dataset labeled with binary relevance, fine-grained scores, and explicit CoT reasoning. An optional GRPO-based optimization step can also be applied to further enhance the model's performance.
During the \emph{inference stage}, inspired by the Qwen3 series~\cite{qwen3}, we employ a special prompt token as the ``think-mode switch'' to force the model to bypass explicit CoT generation and emit only a structured pointwise relevance score, thus minimizing latency and output length, making it suitable for real-time ranking scenarios.

\begin{table}[ht]
    \centering
    \small
    \begin{tabular}{lcc|c}
        \toprule
        \multicolumn{2}{c}{\textbf{Model}} & \textbf{Approach} & \textbf{Avg.} \\
        \midrule
        BM25 & - & Pointwise & 36.9 \\
        Rank1-7B & SFT & Pointwise & 39.2 \\
        Rank1-14B & SFT & Pointwise & 38.7 \\
        REARANK-7B & GRPO & Listwise & 42.8 \\
        Rank-R1-3B & GRPO & Setwise & 37.8 \\
        Rank-R1-7B & GRPO & Setwise & \underline{43.5} \\
        Rank-R1-14B & GRPO & Setwise & \textbf{43.8} \\
        \midrule
        \multicolumn{4}{c}{\textbf{Llama3.2 Series}} \\
        \midrule
        TFRank-1B & SFT & Pointwise & 29.9 \\
        TFRank-3B & SFT & Pointwise & 40.5 \\
        TFRank-8B & SFT & Pointwise & 43.2 \\
        \midrule
        \multicolumn{4}{c}{\textbf{Qwen2.5 Series}} \\
        \midrule
        TFRank-0.5B & SFT & Pointwise & 36.6 \\
        TFRank-1.5B & SFT & Pointwise & 39.5 \\
        TFRank-3B & SFT & Pointwise & 39.7 \\
        TFRank-7B & SFT & Pointwise & 41.5 \\
        \midrule
        \multicolumn{4}{c}{\textbf{Qwen3 Series}} \\
        \midrule
        TFRank-0.6B & SFT & Pointwise & 39.0 \\
        TFRank-1.7B & SFT & Pointwise & 39.4 \\
        TFRank-4B & SFT & Pointwise & 42.5 \\
        TFRank-8B & SFT & Pointwise & 42.1 \\
        \bottomrule
    \end{tabular}
    \caption{
        Short evaluation results (average NDCG@10) on the BEIR benchmark, using BM25 as the retriever.
        TFRank models are trained on each corresponding backbone.
        The best average value for each block is \textbf{bolded}, and the second best is \underline{underlined}.
        Detailed results for all BEIR datasets are provided in the supplementary material.
    }
    \label{tab:beir_short_main_pytrec_eval}
\end{table}

\subsection{Training Data Construction}
High-quality training data is the foundation of TFRank's performance. Our data construction involves two steps:

\subsubsection{Construction of Foundational Relevance Datasets}
To equip the model with fundamental ranking abilities, we first construct two datasets containing binary and fine-grained relevance labels, respectively.
For \textbf{binary relevance data}, following the data processing methodology of Rank1, we acquire approximately 386k positive or negative query-document pairs from the MS MARCO dataset~\cite{msmarco}, using only their binary labels; this dataset is denoted as \texttt{ms\_sub\_binary}.

For \textbf{fine-grained score data}, we follow the methodology of BGE-M3~\cite{bge-m3} to provide the model with more nuanced judgment capabilities. We randomly sample approximately $10\%$ of queries from each MS MARCO length group, along with their top-1 positive and negative documents. For these pairs, we use DeepSeek-R1 to annotate and generate additional documents with fine-grained relevance on a five-point scale. To ensure annotation quality, we apply strict filtering: only positive samples with final scores in $[2, 3, 4]$ and negative samples with scores in $[0, 1]$ are retained. This process yields a fine-grained dataset, denoted as \texttt{ms\_sub\_finegrained}, comprising approximately 7k queries and 44k query-document pairs. 

\subsubsection{Multi-task Data and CoT Distillation}
After constructing the \texttt{ms\_sub\_finegrained} dataset, we expand it into three task paradigms: \textbf{pointwise}, \textbf{pairwise}, and \textbf{listwise}. For pointwise data, we prompt DeepSeek-R1 to generate a CoT explanation for each \texttt{[query, doc]} pair to justify its known fine-grained score, with the final label appended to the end. For the pairwise setting, we sample \texttt{[query, doc$_1$, doc$_2$]} triplets sharing the same query, and use DeepSeek-R1 to generate a CoT explaining their relative relevance and to produce a preference label. For listwise tasks, we follow the REARANK~\cite{zhang2025rearank} format: for each query, we construct a set of documents $D=\{d_1, d_2, ..., d_n\}$, ensuring diversity by sampling no more than two documents per relevance level, and instruct DeepSeek-R1 to generate a CoT that explains the overall ranking logic for the predefined order. This procedure results in a multi-task dataset (\texttt{ms\_sub\_finegrained\_cot}) enriched with CoT reasoning signals across all paradigms.

\begin{figure}[t]
\centering
\includegraphics[width=1.0\columnwidth]{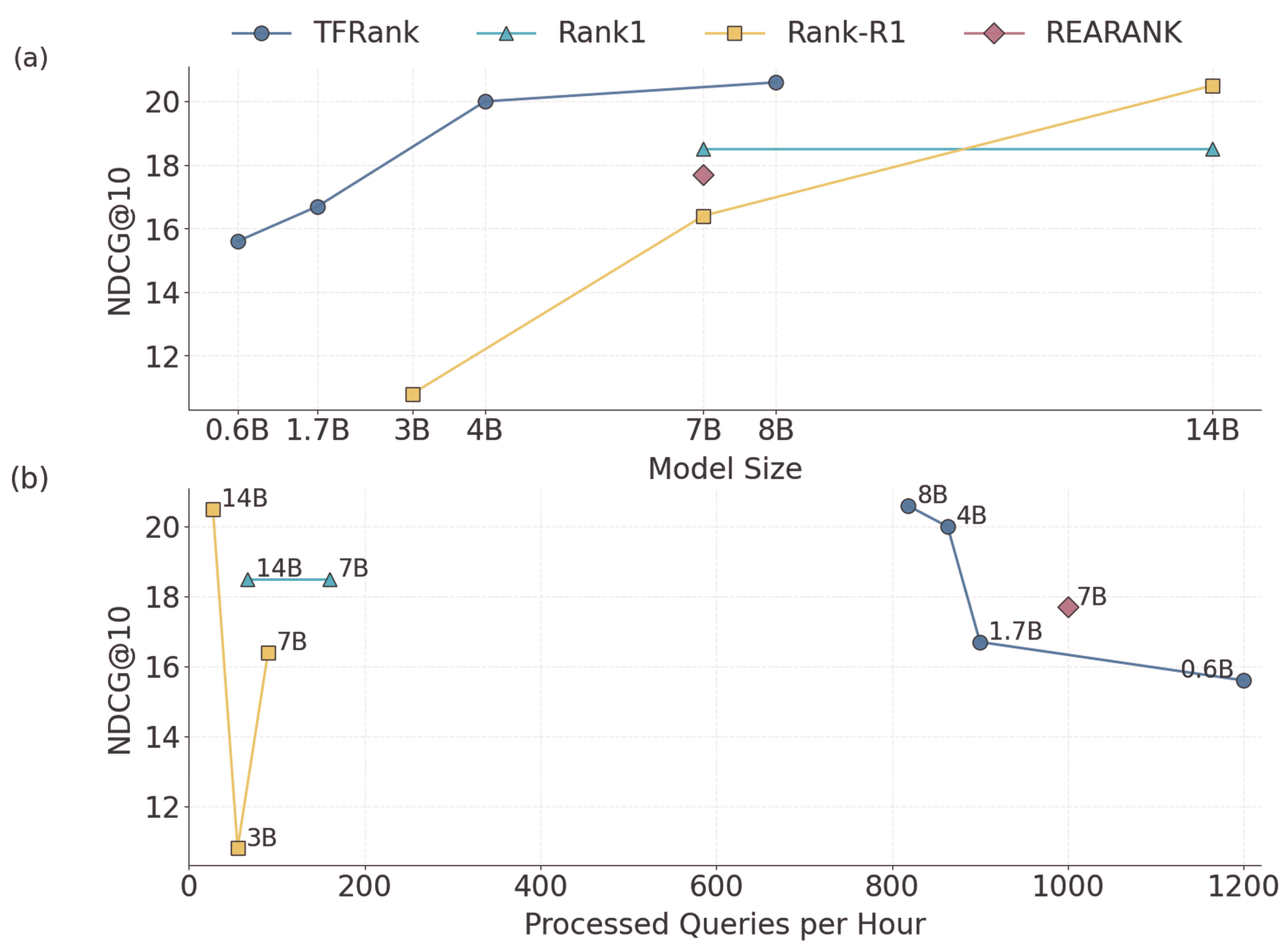} 
\caption{
    Size and efficiency trade-offs for ranking performance on the BRIGHT benchmark. (a) NDCG@10 versus model size for different ranker families; (b) NDCG@10 versus processed queries per hour (efficiency). All TFRank models are trained on the Qwen3 series.
}
\label{fig:size_efficiency_performance}
\end{figure}
\begin{table*}[ht]
    \centering \small
    \setlength{\tabcolsep}{1mm}
    \begin{tabular}{ccc|ccccccc|cc|ccc|c}
        \toprule
        \multirow{2}{*}{\centering\textbf{Model}} & \multirow{2}{*}{\centering\textbf{SFT Setting}} & \multirow{2}{*}{\centering\textbf{Inference Mode}}
        & \multicolumn{7}{c|}{\textbf{StackExchange}} 
        & \multicolumn{2}{c|}{\textbf{Coding}} 
        & \multicolumn{3}{c|}{\textbf{Theorem-based}} 
        & \multirow{2}{*}{\centering\textbf{Avg.}} \\
        \cmidrule(lr){4-10} \cmidrule(lr){11-12} \cmidrule(lr){13-15}
        & & & Bio. & Earth. & Econ. & Psy. & Rob. & Stack. & Sus. 
          & Leet. & Pony 
          & AoPS & TheoQ. & TheoT. 
          & \\
        \midrule
        Qwen3-0.6B & Zero-shot & w/o think & 16.6 & 15.8 & 5.6 & 6.4 & 3.8 & 4.5 & 5.7 & 9.2 & 2.9 & 1.7 & 2.4 & 2.5 & 6.4 \\
        \midrule
        \multirow{5}{*}{TFRank-0.6B} & full & w/o think & 24.8 & 30.0 & 12.0 & 17.5 & 12.9 & 12.1 & 12.7 & 24.4 & 13.1 & 7.1 & 10.3 & 9.8 & 15.6 \\
                        & full & w/ think & 13.9 & 23.2 & 9.9 & 15.1 & 9.5 & 8.8 & 11.7 & 16.4 & 6.8 & 2.6 & 4.8 & 3.9 & 10.5 \\
                        & w/o RR & w/o think & 19.1 & 25.3 & 13.8 & 20.1 & 11.5 & 14.7 & 14.8 & 23.4 & 10.8 & 3.5 & 8.6 & 8.7 & 14.5 \\
                        & w/o FG & w/o think & 25.0 & 26.3 & 12.1 & 20.9 & 13.2 & 10.5 & 15.0 & 25.7 & 5.0 & 6.1 & 9.5 & 11.0 & 15.0 \\
                        & w/o MT & w/o think & 20.5 & 26.6 & 10.2 & 15.8 & 10.4 & 9.3 & 11.0 & 18.5 & 8.5 & 5.6 & 9.3 & 10.3 & 13.0 \\
        \bottomrule
    \end{tabular}
    \caption{
        Ablation results (NDCG@10) of TFRank-0.6B (Qwen3) on BRIGHT using BM25 as the retriever.
        ``full'' is the complete SFT pipeline; ``w/ think'' and ``w/o think'' denote inference with/without explicit reasoning.
        ``w/o RR'' is without Relevance Reasoning; ``w/o FG'' is without Fine-grained Label; ``w/o MT'' is without Multi-task.
    }
    \label{tab:bright_ablation_pytrec_eval}
\end{table*}

\subsection{SFT Training: Internalizing Reasoning}
During the SFT stage, our primary goal is to internalize CoT reasoning capabilities into a small-scale model and teach it to switch between \texttt{/think} and \texttt{/no think} modes based on instructions.

\subsubsection{Think-Mode Switch}
We control the model's generation mode by appending a special token after the input content:
\begin{itemize}
    \item \texttt{/think}: Instructs the model to enter \texttt{/think} mode, requiring it to first generate a reasoning process in the format 
    \texttt{<think>\textbackslash{}n\{reasoning text\}\textbackslash{}n</think>} 
    before providing the final answer.
    \item \texttt{/no think}: Instructs the model to enter \texttt{/no think} mode, bypassing the reasoning process entirely and directly outputting the final answer after an empty \texttt{<think>\textbackslash{}n\textbackslash{}n</think>} tag.
\end{itemize}

\subsubsection{Multi-task Supervised Fine-Tuning}
Based on different task paradigms (listwise, pairwise, and pointwise), constructed training datasets, and the think-mode switch, we construct a unified training data format for SFT:
\begin{itemize}
 \item \texttt{<Instruction><Query><Doc><think mode>}
\end{itemize}
For clarity, we present below an example of a fine-grained \texttt{pointwise} sample in the \texttt{/no think} mode:
\begin{quote}
\textbf{Prompt:} \texttt{<Instruct>}: Please judge the relevance strength between the query and the document, and directly output the relevance judgment (yes or no), followed by the relevance score in parentheses, e.g., yes(score) or no(score). \\
\texttt{<Query>}: what is a stereo preamplifier? \\
\texttt{<Doc>}: Amplifiers are essential components in any sound system, boosting the audio signal to drive loudspeakers and produce audible sound. \\ 
\texttt{/no think} \\
\textbf{Response:} \texttt{<think>\textbackslash{}n\textbackslash{}n</think>no(1)}
\end{quote}

Through this hybrid training strategy, the model not only learns how to perform explicit reasoning but also learns to decouple its reasoning ability from the final judgment and switch its behavior based on instructions. The training loss for the SFT stage is the standard auto-regressive language model cross-entropy loss:
\begin{equation}
\mathcal{L}_{\text{SFT}} = - \sum_{i=1}^{|T|} \log P(t_i | t_{<i}, C)
\label{eq:sft_loss}
\end{equation}
where $C$ is the input prompt and $T = \{t_1, ..., t_{|T|}\}$ is the target sequence.

\subsection{Inference: ``Think-Free'' Pointwise Ranking}
During the inference stage, TFRank's design is entirely guided by efficiency and practicality.

\subsubsection{Activating \texttt{/no think} Mode} For all inference requests, we uniformly append the \texttt{/no think} command to the prompt. This forces the model to suppress explicit CoT generation, causing it to output only a short answer containing the relevance judgment and score, such as \texttt{yes(4)} or \texttt{no(1)}. This approach reduces the generation length from hundreds of tokens to just a few, achieving an order-of-magnitude reduction in latency.

\subsubsection{Pointwise Score Extraction and Fusion} The model's raw output contains two dimensions of signals: a binary judgment (yes/no) and a five-class fine-grained score (0-4). To obtain an expressive final ranking score, we designed a fusion formula.
First, we calculate the probability of the binary judgment being ``yes'', $P_{bi}$, from the model's output logits:
    \begin{equation}
    P_{bi} = \text{softmax}([\text{logits}_{\text{yes}}, \text{logits}_{\text{no}}])[0]
    \end{equation}
    where $\text{logits}_{\text{yes}}$ and $\text{logits}_{\text{no}}$ are the model's predicted logits for the ``yes'' and ``no'' tokens, respectively.

    Next, we compute the expected value of the fine-grained score $S_{fg}$:
    \begin{equation}
    S_{fg} = \frac{\sum_{i=0}^{4} i \cdot \text{softmax}(\text{logits}_{0...4})}{\max(i) - \min(i)}
    \end{equation}
    where $\text{logits}_{0...4}$ are the model's predicted logits for the five score tokens ``0'' through ``4''.

    Finally, we average these two scores to obtain the final TFRank sorting score:
    \begin{equation}
    \text{Score}_{\text{TFRank}} = 0.5 \cdot P_{bi} + 0.5 \cdot S_{fg}
    \end{equation}
\begin{figure}[t]
\centering
\includegraphics[width=1.0\columnwidth]{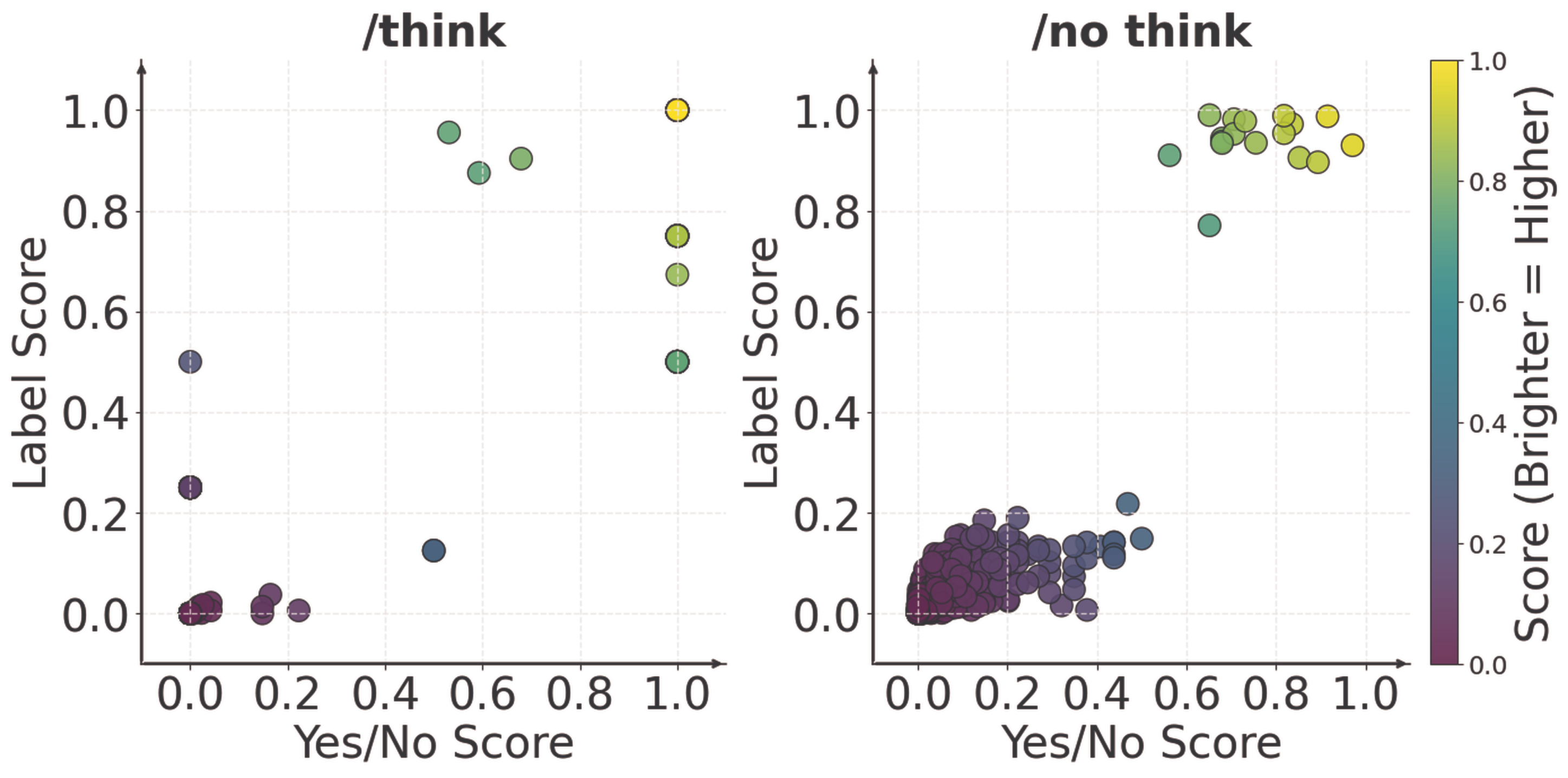} 
\caption{
    Score distributions for a random 1\% sample of BRIGHT, evaluated by TFRank-0.6B (Qwen3) under \texttt{/think} and \texttt{/no think} inference modes.
}
\label{fig:think_nothink_distribution}
\end{figure}
\begin{table*}[ht]
    \centering \small
    \setlength{\tabcolsep}{1.5mm}
    \begin{tabular}{cc|ccccccc|cc|ccc|c}
        \toprule
        \multicolumn{2}{c}{\multirow{2}{*}{\centering\textbf{Model}}} 
        & \multicolumn{7}{|c|}{\textbf{StackExchange}} 
        & \multicolumn{2}{c|}{\textbf{Coding}} 
        & \multicolumn{3}{c|}{\textbf{Theorem-based}} 
        & \multirow{2}{*}{\centering\textbf{Avg.}} \\
        \cmidrule(lr){3-9} \cmidrule(lr){10-11} \cmidrule(lr){12-14}
        & & Bio. & Earth. & Econ. & Psy. & Rob. & Stack. & Sus. 
          & Leet. & Pony 
          & AoPS & TheoQ. & TheoT. 
          & \\
        \midrule
        \multirow{5}{*}{TFRank-0.6B} 
                                     & SFT  & 24.8 & 30.0 & 12.0 & 17.5 & 12.9 & 12.1 & 12.7 & 24.4 & 13.1 & 7.1 & 10.3 & 9.8 & 15.6 \\
                                     & GRPO  & 14.9 & 25.8 & 10.1 & 13.9 & 16.0 & 11.9 & 13.1 & 20.6 & 8.3 & 4.1 & 7.5 & 4.7 & 12.6 \\
                                     & SFT+GRPO & 23.1 & 28.0 & 15.1 & 18.2 & 11.9 & 13.8 & 13.9 & 23.8 & 10.6 & 6.8 & 8.6 & 9.9 & 15.3 \\
                                     & GRPO$^{\ddagger}$ & 32.1 & 43.8 & 16.4 & 22.7 & 23.2 & 16.9 & 19.1 & 25.0 & 12.9 & 5.1 & 8.2 & 5.8 & 19.3 \\
                                     & SFT+GRPO$^{\ddagger}$ & 21.7 & 33.5 & 17.0 & 25.9 & 14.6 & 17.2 & 17.1 & 26.1 & 10.1 & 6.7 & 9.0 & 9.4 & 17.4 \\ \midrule
        \multirow{3}{*}{TFRank-1.7B} 
                                     & SFT  & 25.2 & 29.7 & 17.2 & 26.2 & 15.0 & 16.7 & 17.9 & 21.9 & 10.1 & 4.5 & 7.0 & 9.4 & 16.7 \\
                                     & SFT+GRPO & 24.9 & 29.5 & 16.4 & 26.2 & 14.6 & 15.6 & 17.5 & 27.5 & 9.0 & 6.5 & 10.1 & 9.6 & 17.3 \\
                                     & GRPO$^{\ddagger}$ & 30.9 & 42.1 & 19.1 & 25.4 & 14.8 & 21.9 & 18.8 & 19.5 & 15.3 & 5.3 & 8.5 & 10.3 & 19.3 \\ \midrule
        \multirow{3}{*}{TFRank-4B} 
                                     & SFT  & 31.4 & 40.9 & 19.4 & 26.2 & 18.8 & 19.1 & 20.3 & 23.4 & 13.0 & 7.7 & 10.1 & 9.1 & 20.0 \\
                                     & SFT+GRPO & 24.8 & 34.9 & 16.2 & 21.2 & 19.0 & 19.4 & 15.7 & 22.0 & 11.2 & 5.1 & 10.1 & 9.1 & 17.4 \\
                                     & GRPO$^{\ddagger}$ & 35.3 & 45.8 & 21.4 & 31.1 & 19.8 & 24.3 & 24.9 & 26.2 & 16.7 & 6.8 & 9.3 & 11.6 & 22.8 \\ \midrule
        \multirow{3}{*}{TFRank-8B} 
                                     & SFT  & 29.8 & 42.3 & 21.5 & 25.9 & 19.7 & 21.3 & 22.8 & 21.6 & 16.4 & 6.8 & 10.4 & 9.0 & 20.6 \\
                                     & SFT+GRPO & 30.3 & 40.6 & 21.4 & 26.1 & 19.1 & 20.2 & 22.5 & 29.1 & 14.9 & 9.8 & 10.9 & 10.1 & 21.3 \\
                                     & GRPO$^{\ddagger}$ & 34.5 & 48.6 & 23.7 & 28.3 & 21.9 & 22.3 & 26.8 & 28.1 & 21.2 & 7.2 & 9.5 & 11.0 & 23.6 \\
        \bottomrule
    \end{tabular}
    \caption{
        Results (NDCG@10) of TFRank (Qwen3) on BRIGHT under different training paradigms, using BM25 as the retriever. Models marked with $^{\ddagger}$ apply GRPO using all training samples, while unmarked models use GRPO on a randomly sampled approximately 20\% subset of queries for efficiency.}
    \label{tab:bright_yesno_grpo_pytrec_eval}
\end{table*}
\subsection{Optional: Optimization with GRPO}
To further explore the upper bound of relevance modeling in TFRank, we employ GRPO as an advanced optimization strategy. We compare two paradigms: \textbf{SFT-then-GRPO}, where the model first undergoes multi-task SFT before further GRPO optimization (labeled as \texttt{SFT+GRPO} in Table~\ref{tab:bright_yesno_grpo_pytrec_eval}); and \textbf{Direct GRPO}, where end-to-end GRPO is directly applied to the instruction-tuned base model. 

We design a multi-dimensional reward function $R_{\text{total}}$ to optimize both ranking quality and output consistency:
\begin{equation}
    R_{\text{total}} = R_{\text{content}} + \lambda\, R_{\text{format}}
\end{equation}
where $\lambda$ is a balancing hyperparameter (default: $0.5$).

\noindent
\textbf{Content Reward ($R_{\text{content}}$):}
The content reward is tailored for each task paradigm:
for \emph{pointwise} tasks, it averages binary classification accuracy and the mean squared error (MSE) of relevance scores;
for \emph{pairwise}, it averages preference accuracy and the MSE scores;
For \emph{listwise}, it is the mean of ranking relation accuracy, MSE scores, and relative nDCG improvement (as defined in REARANK).

\noindent
\textbf{Format Reward ($R_{\text{format}}$):}
The format reward enforces strict output conventions, including proper use of the \texttt{<think>} tag, minimum reasoning length in \texttt{/think} mode, and adherence to the answer format required by each paradigm (e.g., \texttt{yes/no(score)} for pointwise).

\section{Experiments}

\paragraph{Datasets and Metrics}
We evaluate TFRank on two representative benchmarks: BRIGHT~\cite{bright} and BEIR~\cite{beir}. BRIGHT contains approximately 1,384 real-world queries from diverse domains such as coding, mathematics, and economics, specifically designed to assess multi-step reasoning capabilities beyond keyword matching. BEIR comprises 18 public datasets covering a wide variety of retrieval tasks, including question answering, argument retrieval, fact-checking, biomedical search, and duplicate question detection, emphasizing zero-shot generalization across unseen domains. Following prior work, we primarily report NDCG@10, which effectively measures ranking quality at top positions.

\paragraph{Baselines}
We compare TFRank with several representative ranking methods. BM25 is included as a classical lexical baseline. We consider zero-shot LLM-based listwise ranking, such as RankGPT~\cite{rankgpt}, and recent state-of-the-art methods including Rank1~\cite{weller2025rank1}, Rank-R1~\cite{zhuang2025rank}, and REARANK~\cite{zhang2025rearank}. All baselines are evaluated using official implementations or released checkpoints.

\paragraph{Implementation Details}
TFRank models are trained on three LLM families: Qwen2.5 (0.5B$ \sim $7B)~\cite{qwen2.5}, Qwen3 (0.6B$ \sim $8B)~\cite{qwen3}, and Llama 3.2 (1B$ \sim $8B)~\cite{llama3}. Llama-3.2-8B is sourced from the open-source community\footnote{\url{https://modelscope.cn/models/voidful/Llama-3.2-8B-Instruct}}
, and the rest from official repositories. 
Our training setup adopts a learning rate of $1\times10^{-5}$ with random seed 42, and uses at most 5 SFT epochs. In the GRPO training stage, each instance produces eight completions. For checkpoint selection, $1\%$ of the training data is reserved as a validation set. All experiments are performed on one 8-GPU H20 server. More detailed prompts, training data statistics, and full reward definitions are provided in the Appendix and the released code.
\section{Results and Analysis}
\subsection{Main Results}
\paragraph{Overall Performance}
As illustrated in Table~\ref{tab:bright_yesno_main_pytrec_eval}, on the reasoning-intensive BRIGHT benchmark, TFRank \textbf{consistently and significantly outperforms} all existing baselines across multiple model families. Notably, TFRank unleashes the potential of small-scale LLMs. For example, Qwen3-based SFT TFRank-1.7B achieves an average NDCG@10 of 16.7, competing with the \textbf{4×}-larger Rank1-7B (18.5), REARANK-7B (17.7), and Rank-R1-7B (16.4). Similarly, TFRank-3B, based on Llama-3.2 and Qwen-2.5, even outperforms some aforementioned 7B baselines.
On the general-purpose BEIR benchmark, in Table~\ref{tab:beir_short_main_pytrec_eval}, TFRank delivers performance on par with the strongest baselines. For instance, TFRank-8B, based on Llama-3.2, achieves an average NDCG@10 of 43.2, outperforming Rank1-14B and REARANK-7B, while remaining comparable to the 43.5 and 43.8 achieved by Rank-R1‐7B and Rank-R1‐14B.

\paragraph{Model Efficiency}
While achieving high performance, TFRank demonstrates \textbf{significant efficiency advantages}. As shown in Figure~\ref{fig:size_efficiency_performance}(a), on the BRIGHT benchmark, the TFRank performance curve is consistently above other models, indicating its ability to achieve superior performance at equivalent or even smaller model scales, which is crucial for resource-constrained deployment scenarios.
Figure~\ref{fig:size_efficiency_performance}(b) further reveals its inference efficiency. By adopting a pointwise scoring and ``Think‐Free'' reasoning mode, TFRank processes far more queries per hour than methods that require explicit CoT generation, such as Rank1 and Rank‐R1. 
Compared to the listwise model REARANK, TFRank's pointwise scoring enables intra-query parallelism, allowing single-query throughput to scale effectively with hardware resources. This is particularly advantageous for latency-sensitive applications. Remarkably, the 0.6B TFRank surpasses the 7B REARANK in performance on BRIGHT, while offering a significant throughput advantage.

\begin{figure}[t]
    \centering
    \includegraphics[width=1.0\columnwidth]{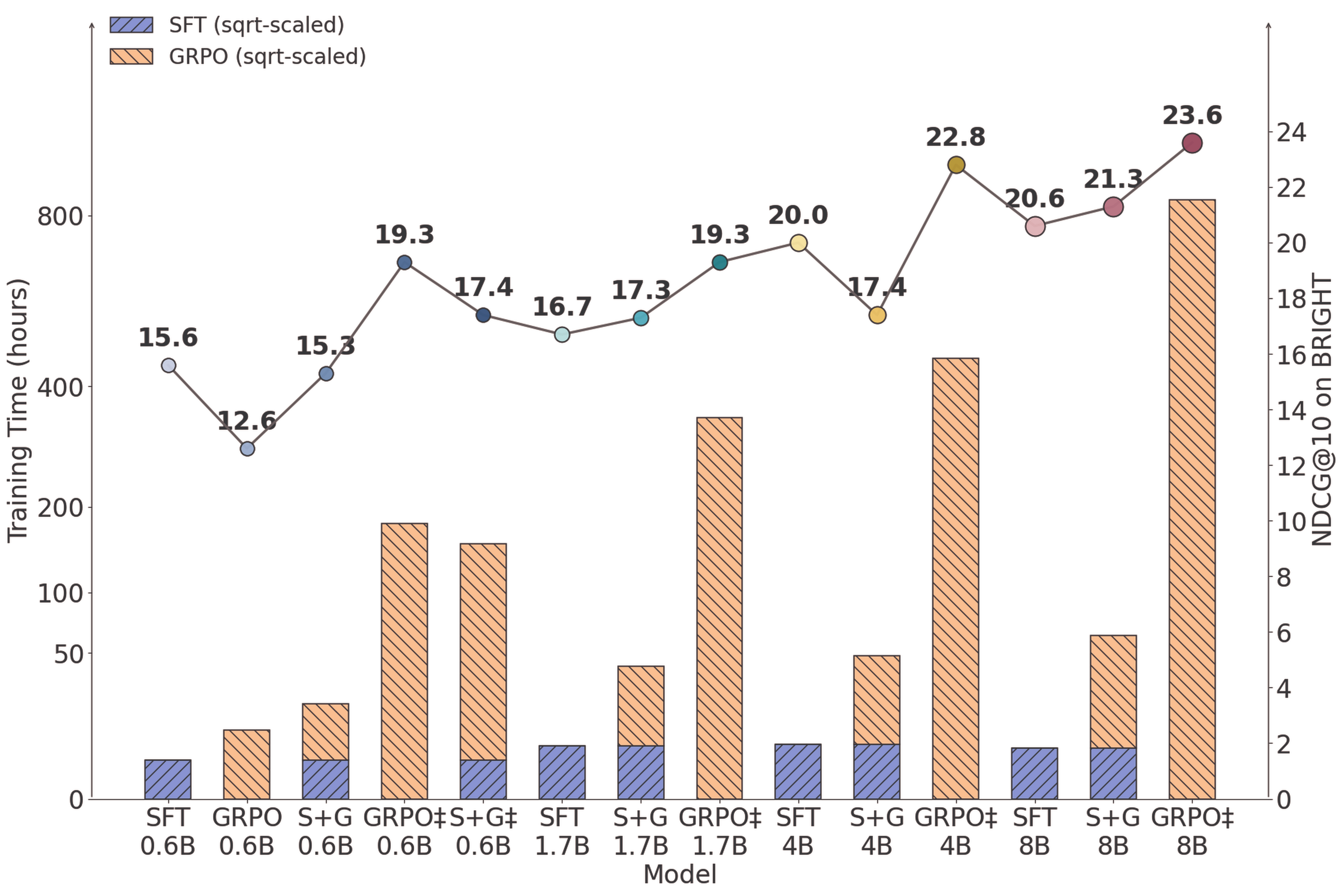} 
    \caption{
    Training time (hours) versus ranking performance (NDCG@10 on BRIGHT) for different training strategies. All TFRank models use the Qwen3 series backbone. The meaning of the $^{\ddagger}$ symbol follows that in Table~\ref{tab:bright_yesno_grpo_pytrec_eval}.
    }
    \label{fig:training_time_performance}
\end{figure}
\subsection{Ablation Study}
We conducted a series of ablation studies on the TFRank-0.6B (Qwen3) model, with the results presented in Table~\ref{tab:bright_ablation_pytrec_eval}.
\paragraph{Effectiveness of Components}
Our full TFRank model achieves a performance of 15.6, more than doubling the score attained by its base model and demonstrating the overall effectiveness of our training method. Ablation experiments confirm that each of our proposed components is indispensable: removing either Relevance Reasoning Supervision, Fine-grained Relevance Labels, or Multi-task training leads to a clear performance decline.

\paragraph{``Think-Free'' Reasoning}
A striking finding is the effectiveness of ``Think-Free'' reasoning. Forcing explicit CoT at inference (``w/ think'') severely damages performance, reducing NDCG@10 to 10.5.  As illustrated in Figure~\ref{fig:think_nothink_distribution}, we observe that the score distribution in the \texttt{/think} mode is more concentrated, indicating a weaker capacity to discern documents with partial relevance. 
This aligns with observations by~\citet{overthink}: explicit reasoning can induce overconfidence and impair the model's ability to model partial relevance finely.
By internalizing the reasoning process during training, TFRank can output more precise and discriminative scores at inference time without explicit ``thinking'', thereby achieving a substantial increase in efficiency.

\subsection{Further Optimization with GRPO}
To explore the performance ceiling of TFRank, we introduced the GRPO training strategy.
As shown in Table~\ref{tab:bright_yesno_grpo_pytrec_eval}, 
1.7B and 8B models show small but consistent gains when GRPO is applied after SFT. Interestingly, across the full Qwen3 series, directly applying full-data GRPO yields significant improvements over SFT or SFT+GRPO. This may be because Qwen3 models, as reasoning-oriented LLMs, can better exploit GRPO’s broader policy exploration to discover more optimal reasoning and ranking strategies than standard SFT.
However, this performance improvement comes at a significant training cost. As depicted in Figure~\ref{fig:training_time_performance}, the training duration for GRPO far exceeds that of SFT. For example, the cost of full GRPO training on the 0.6B model (approx. 200 hours) is much higher than that of training a superior-performing 8B SFT model (approx. 8 hours).
Therefore, while increasing model scale via SFT or increasing the amount of GRPO data can both effectively boost performance, they represent a trade-off between computational resources and time.

\subsection{BRIGHT Leaderboard and Domain Generalization}
To validate TFRank’s competitiveness within state-of-the-art retrieval systems, we employ ReasonIR-8B~\cite{reasonir} as the retriever and compare both our models and baselines against top BRIGHT leaderboard entries. To further examine TFRank’s generalization ability on reasoning-intensive ranking, we evaluate it on a domain-specific dataset, R2MED~\cite{li2025r2med}, which focuses on complex medical retrieval, using E5-mistral-7b-instruct~\cite{wang2023improving} as the retriever. Detailed results are provided in the Appendix. Experimental results show that GRPO TFRank-8B achieves an NDCG@10 of 32.0 on BRIGHT, substantially outperforming REARANK-7B (24.2) and Rank-R1-7B (23.2), while remaining slightly behind much larger models such as Rank-R1-32B. On R2Med, TFRank reaches an NDCG@10 of 34.7, surpassing strong baselines such as Rank-R1-7B (32.87) and Rank1-7B (32.30). These results demonstrate that TFRank is a competitive and generalizable alternative when deploying ultra-large models is not feasible.

\section{Conclusion}
We introduce TFRank, an innovative framework that addresses the efficiency bottlenecks of reasoning-intensive rankers in real-world deployments. Through a “Think-Free” reasoning mechanism, TFRank uses multi-task learning and CoT supervision during training to internalize complex reasoning capabilities within small-scale LLMs. During inference, it completely bypasses the generation of explicit reasoning chains, directly and efficiently producing pointwise relevance scores.
Experiments demonstrate that this approach enables small-scale models (e.g., 1.7B) to compete with baselines with over four times the parameters, while substantially reducing inference latency. We further observe that once reasoning is internalized, suppressing CoT generation during inference yields more accurate ranking results.
In summary, TFRank provides a practical path for building and deploying advanced reasoning-based rankers, offering a feasible solution to popularize complex AI capabilities in resource-constrained applications.

\section{Acknowledgments}
We would like to express our sincere gratitude to the anonymous reviewers for their valuable feedback. We also thank the Chairs and the organizing staff for their dedicated efforts in facilitating this work. This work is supported by the National Natural Science Foundation of China (62272439/62572456) and the Fundamental Research Funds for the Central Universities.

\bibliography{aaai2026}

\clearpage
\appendix 

\section{Supplementary Material Overview}
This appendix provides additional details and resources that complement the main paper. We include (i) extended descriptions and statistics of the datasets used, (ii) some comprehensive experimental results, and further notes on inference settings and implementation. 

\renewcommand{\thetable}{A\arabic{table}}
\renewcommand{\thefigure}{A\arabic{figure}}
\setcounter{figure}{0}
\setcounter{table}{0}
\section{Dataset Supplementary Material}
\subsection{Dataset Statistics}
Table~\ref{tab:train_data_stats} presents the detailed statistics of each training and evaluation data subset, including the number of queries, positive/negative pairs, and fine-grained annotations. Further breakdowns by task paradigm (pointwise, pairwise, listwise) are provided.

\begin{table*}[ht]
    \centering
    \small
    \begin{tabular}{llcccccc}
        \toprule
        \textbf{Dataset} & \textbf{Source} & \textbf{Task} & \textbf{RR} & \textbf{FG} & \textbf{\#Queries} & \textbf{\#Samples} \\
        \midrule
        ms\_sub\_binary & MS MARCO, Rank1 & Pointwise & N & N & 195,335 & 385,791 \\
        \midrule
        \multirow{3}{*}{ms\_sub\_finegrained}
            & \multirow{3}{*}{MS MARCO, M3 data, Deepseek-R1}
            & Pointwise  & Y & Y & 7,068  & 44,605 $\times$ 2 $\times$ 2 \\
            &           & Pairwise   & Y & Y & 7,068   & 57,351 $\times$ 2 $\times$ 2  \\
            &           & Listwise   & Y & Y & 7,068   & 13,852 $\times$ 2 $\times$ 2  \\
        \bottomrule
    \end{tabular}
    \caption{
        Statistics of training data used in TFRank. ``RR'' indicates whether Relevance Reasoning (\texttt{/think} mode) is included (Y) or not (\texttt{/no think} mode, N); ``FG'' denotes the availability of fine-grained labels; ``\#'' refers to the number of queries or samples. Multiplication terms (e.g., $\,\times2\times2$) indicate duplication arising from both RR and FG settings.
    }
    \label{tab:train_data_stats}
\end{table*}

\subsection{Prompt Formats}
\begin{figure*}[ht]
\centering
\includegraphics[width=0.75\linewidth]{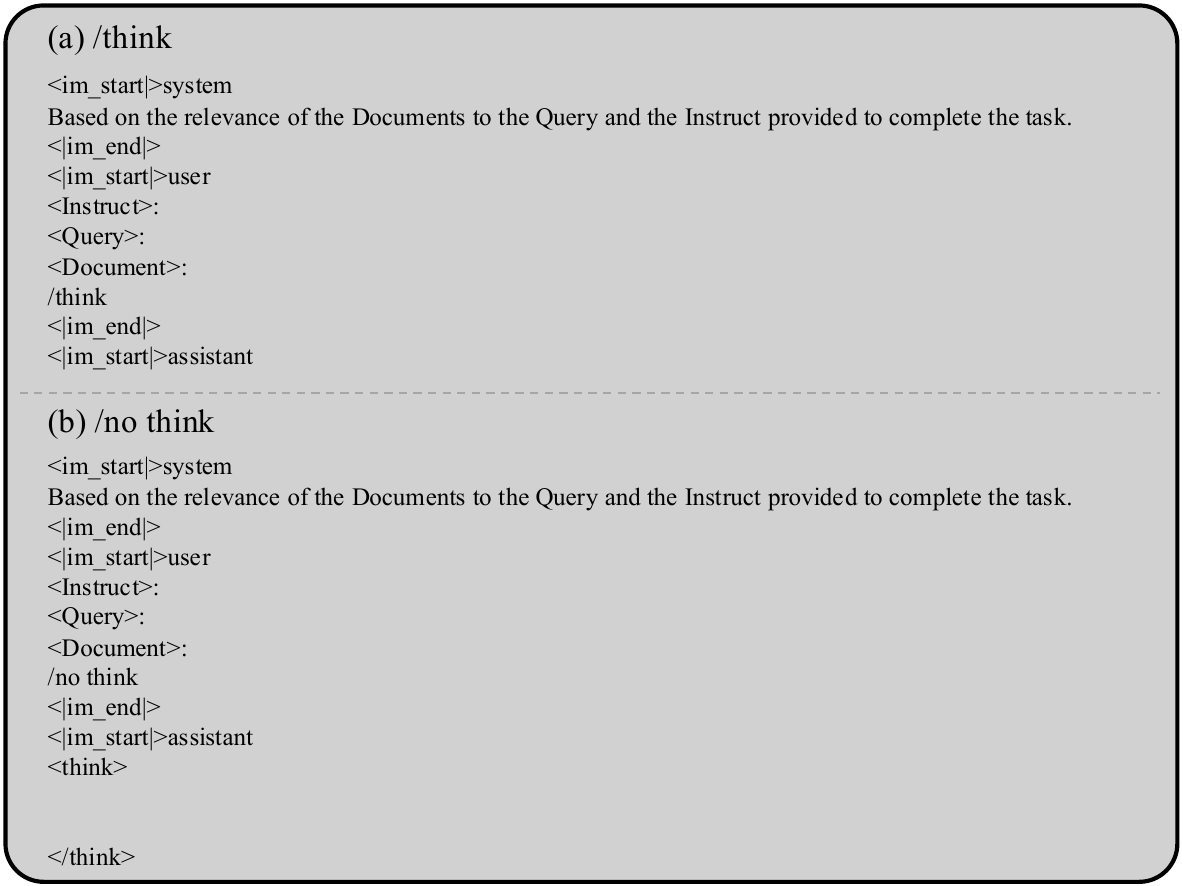} 
\caption{
Prompt templates for TFRank inference. (a) depicts the completion prompt for the \texttt{/think} mode, which requires the model to output an explicit reasoning process. (b) shows the template for the \texttt{/no think} mode, instructing the model to return a relevance score without reasoning directly.
}
\label{fig:inference_template}
\end{figure*}

\begin{figure*}[ht]
\centering
\includegraphics[width=0.75\linewidth]{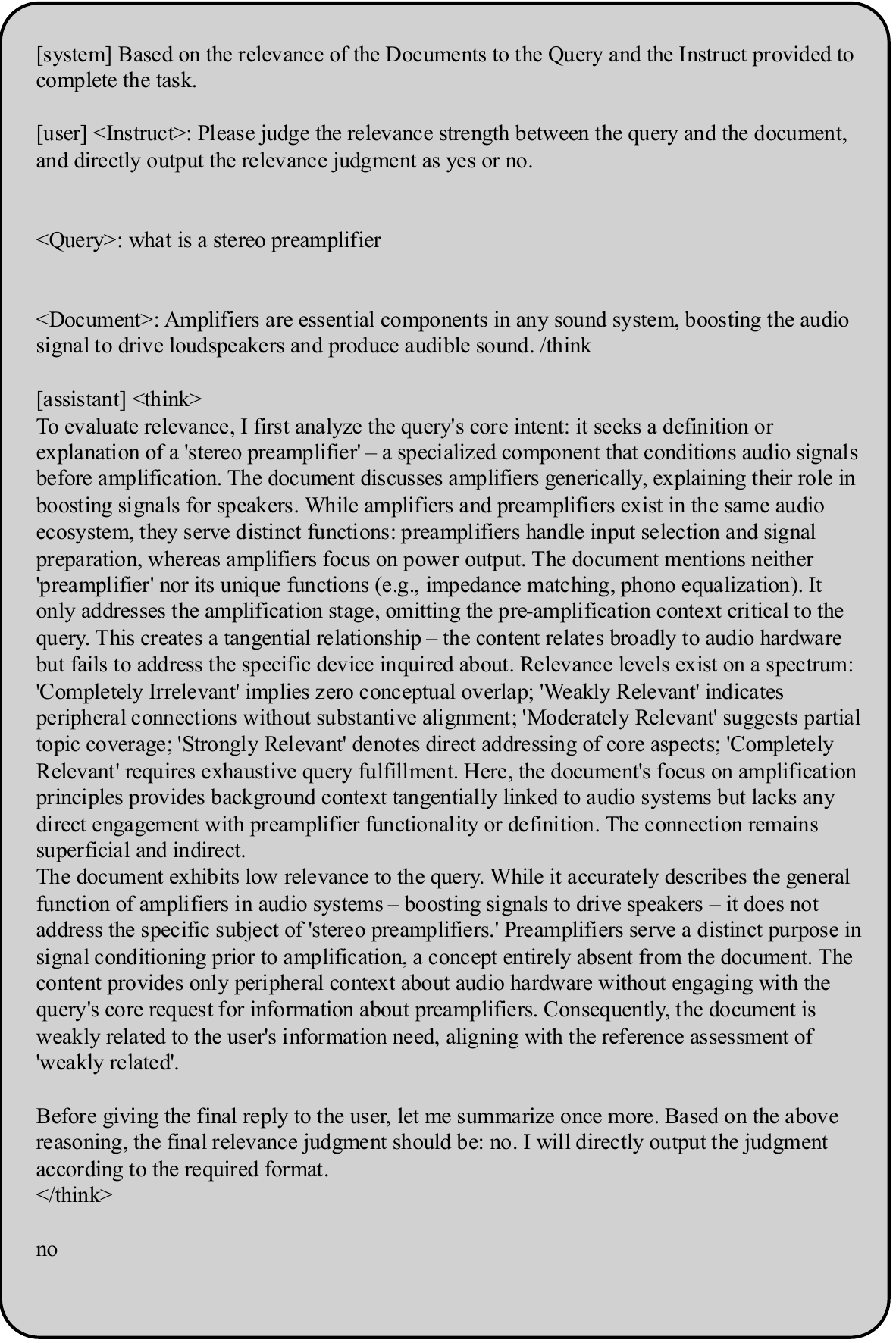} 
\caption{
    Example of a pointwise binary (\texttt{yes/no}) training sample in \texttt{messages} format with explicit reasoning (\texttt{/think} mode) in TFRank.
}
\label{fig:pointwise_binary_think}
\end{figure*}

\begin{figure*}[ht]
\centering
\includegraphics[width=0.75\linewidth]{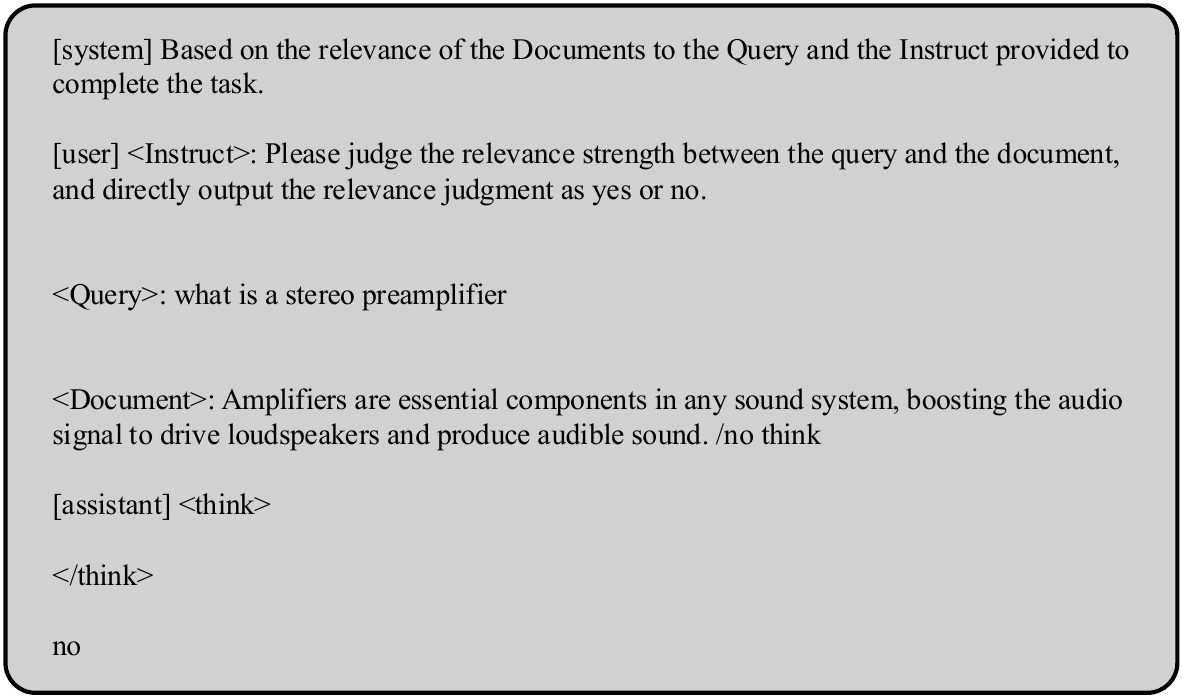} 
\caption{
    Example of a pointwise binary (\texttt{yes/no}) training sample in \texttt{messages} format without explicit reasoning (\texttt{/no think} mode) in TFRank.
}
\label{fig:pointwise_binary_no_think}
\end{figure*}

\begin{figure*}[ht]
\centering
\includegraphics[width=0.75\linewidth]{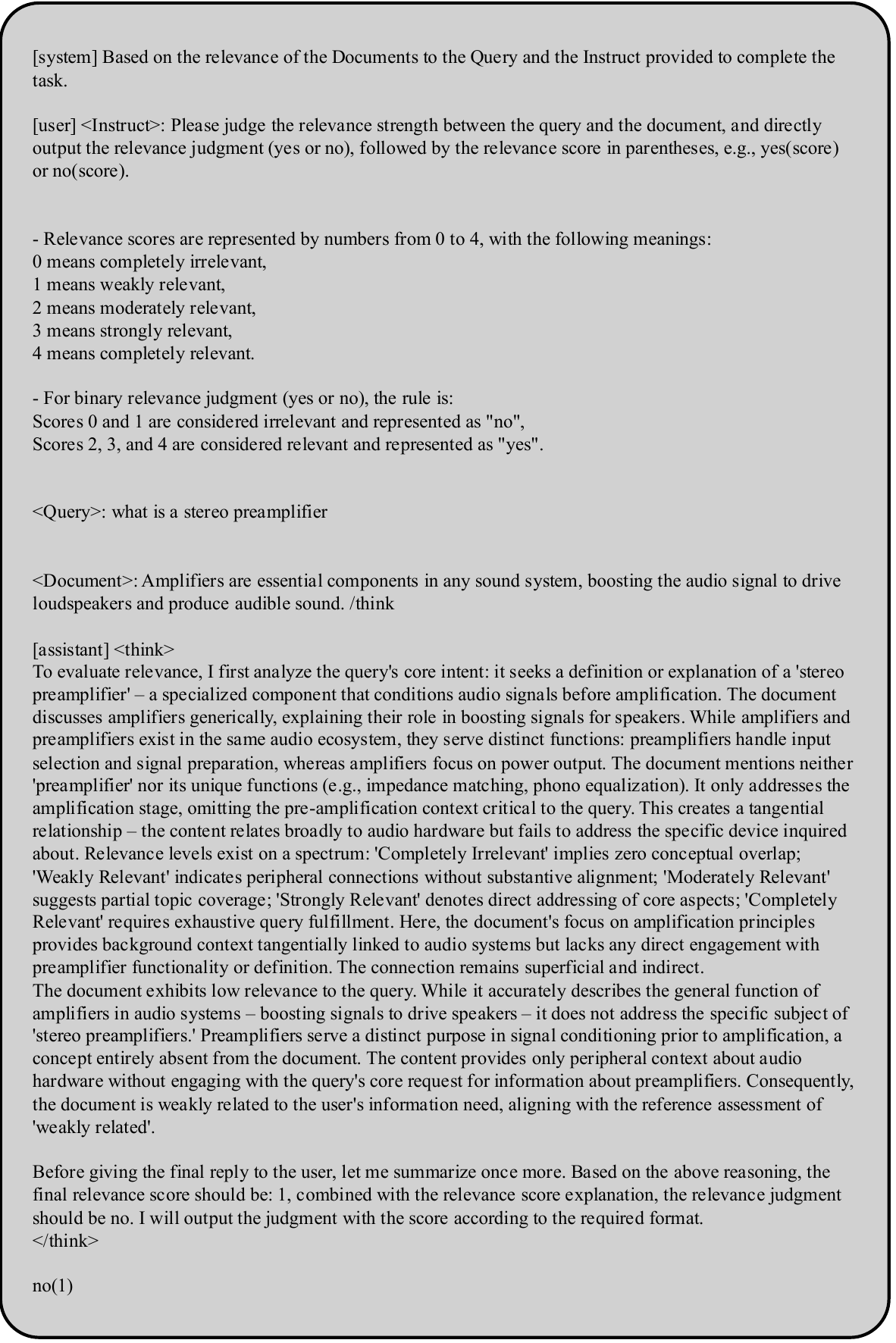} 
\caption{
    Example of a pointwise fine-grained training sample in \texttt{messages} format with explicit reasoning (\texttt{/think} mode) in TFRank.
}
\label{fig:pointwise_fg_think}
\end{figure*}

\begin{figure*}[ht]
\centering
\includegraphics[width=0.75\linewidth]{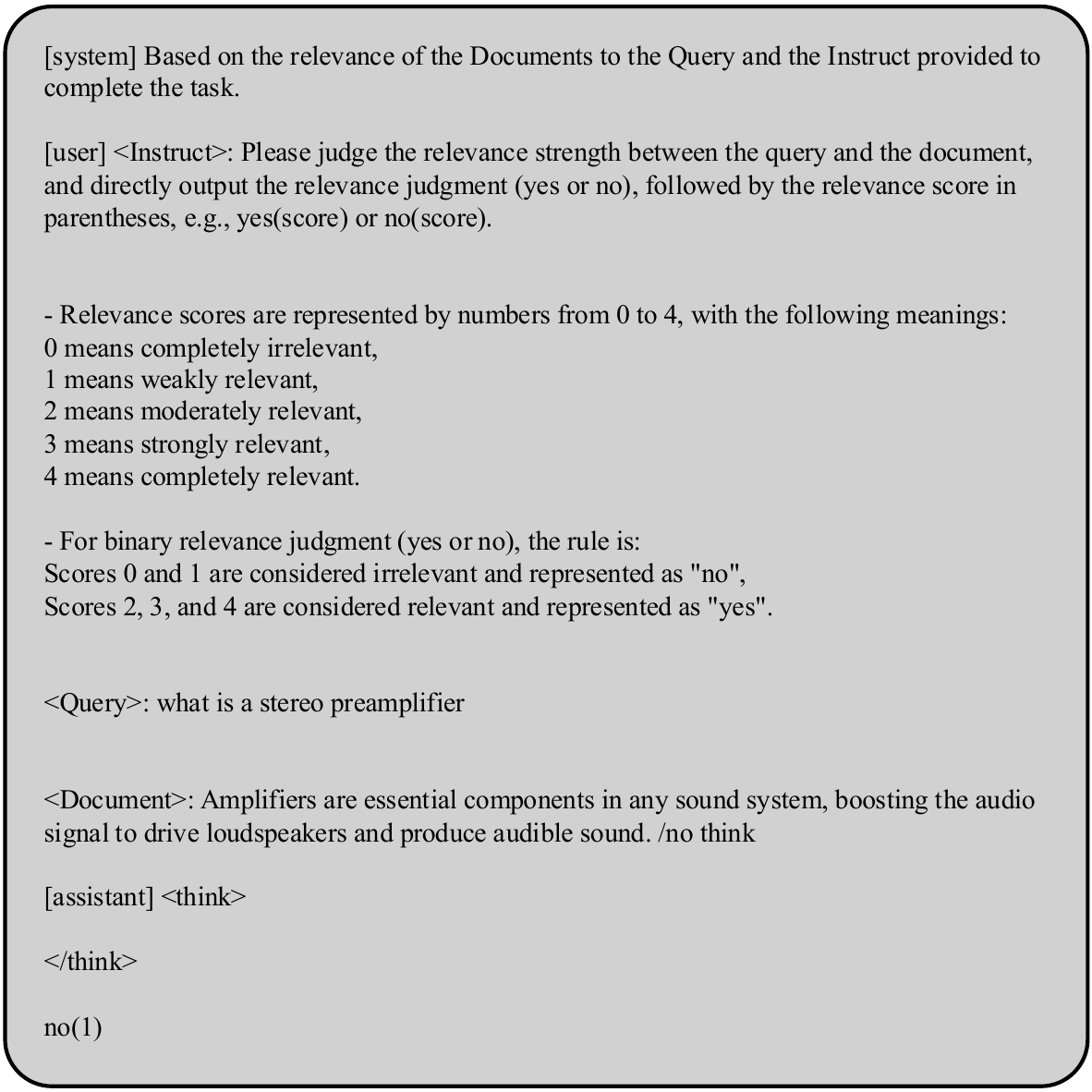} 
\caption{
    Example of a pointwise fine-grained training sample in \texttt{messages} format without explicit reasoning (\texttt{/no think} mode) in TFRank.
}
\label{fig:pointwise_fg_no_think}
\end{figure*}

\begin{figure*}[ht]
\centering
\includegraphics[width=0.75\linewidth]{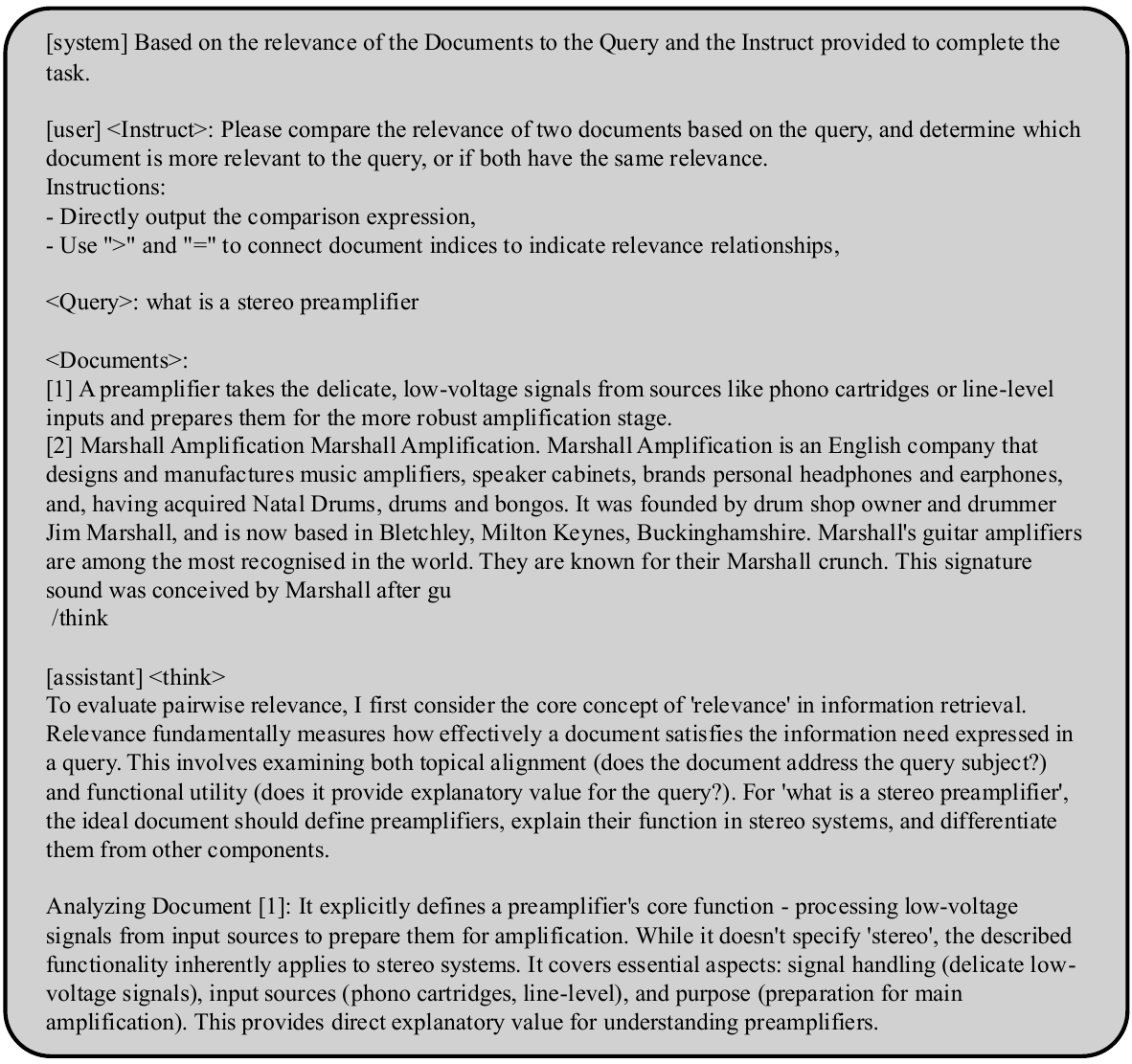} 
\caption{
    Example of a pairwise ID-only training sample in \texttt{messages} format with explicit reasoning (\texttt{/think} mode) in TFRank (1/2).
}
\label{fig:pairwise_id_think_1}
\end{figure*}

\begin{figure*}[ht]
\centering
\includegraphics[width=0.75\linewidth]{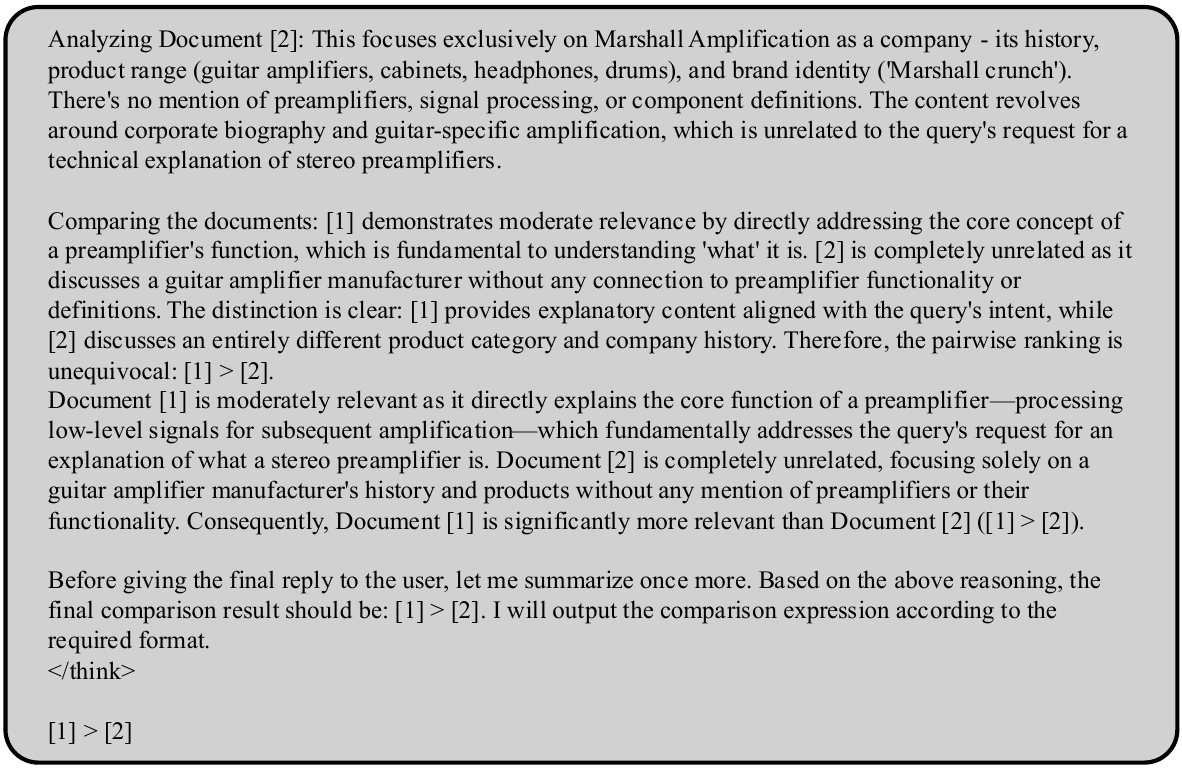} 
\caption{
    Example of a pairwise ID-only training sample in \texttt{messages} format with explicit reasoning (\texttt{/think} mode) in TFRank (2/2).
}
\label{fig:pairwise_id_think_2}
\end{figure*}

\begin{figure*}[ht]
\centering
\includegraphics[width=0.75\linewidth]{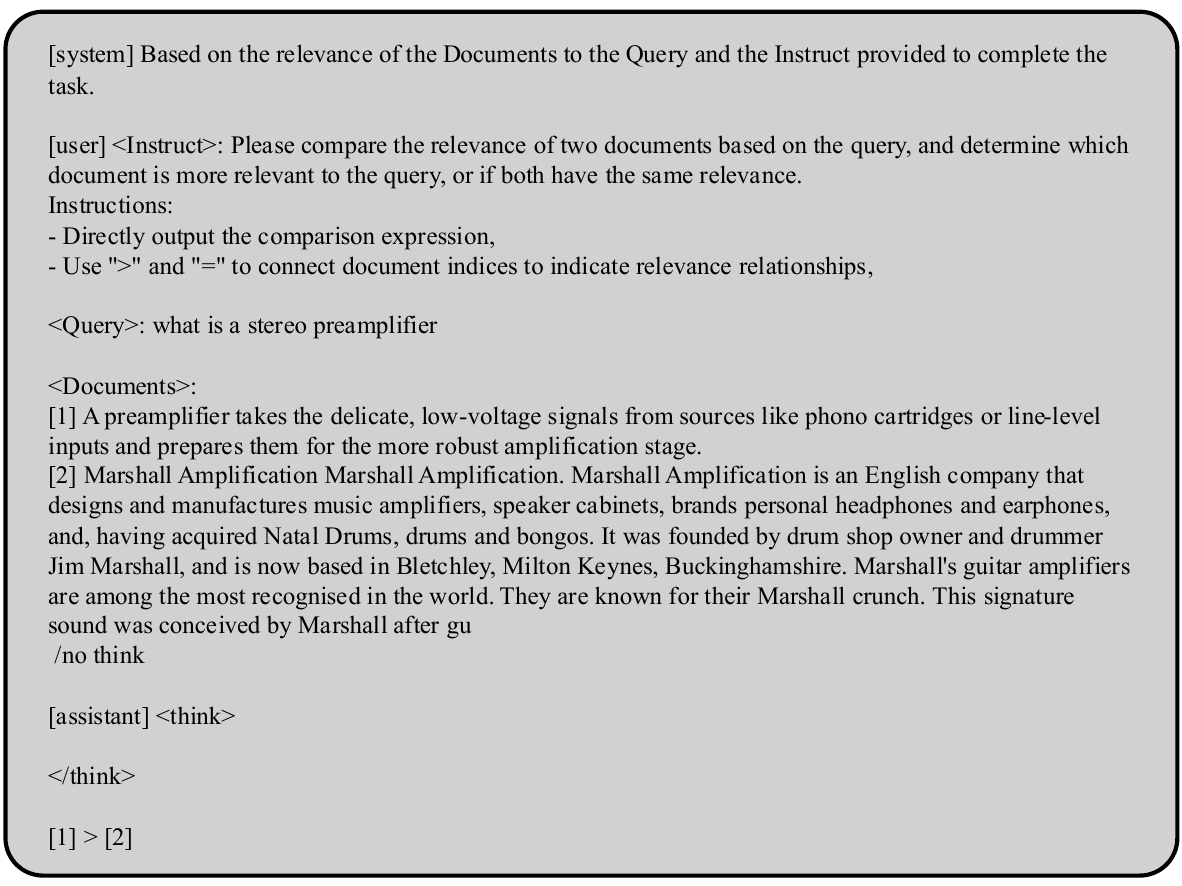} 
\caption{
    Example of a pairwise ID-only training sample in \texttt{messages} format without explicit reasoning (\texttt{/no think} mode) in TFRank.
}
\label{fig:pairwise_id_nothink}
\end{figure*}

\begin{figure*}[ht]
\centering
\includegraphics[width=0.75\linewidth]{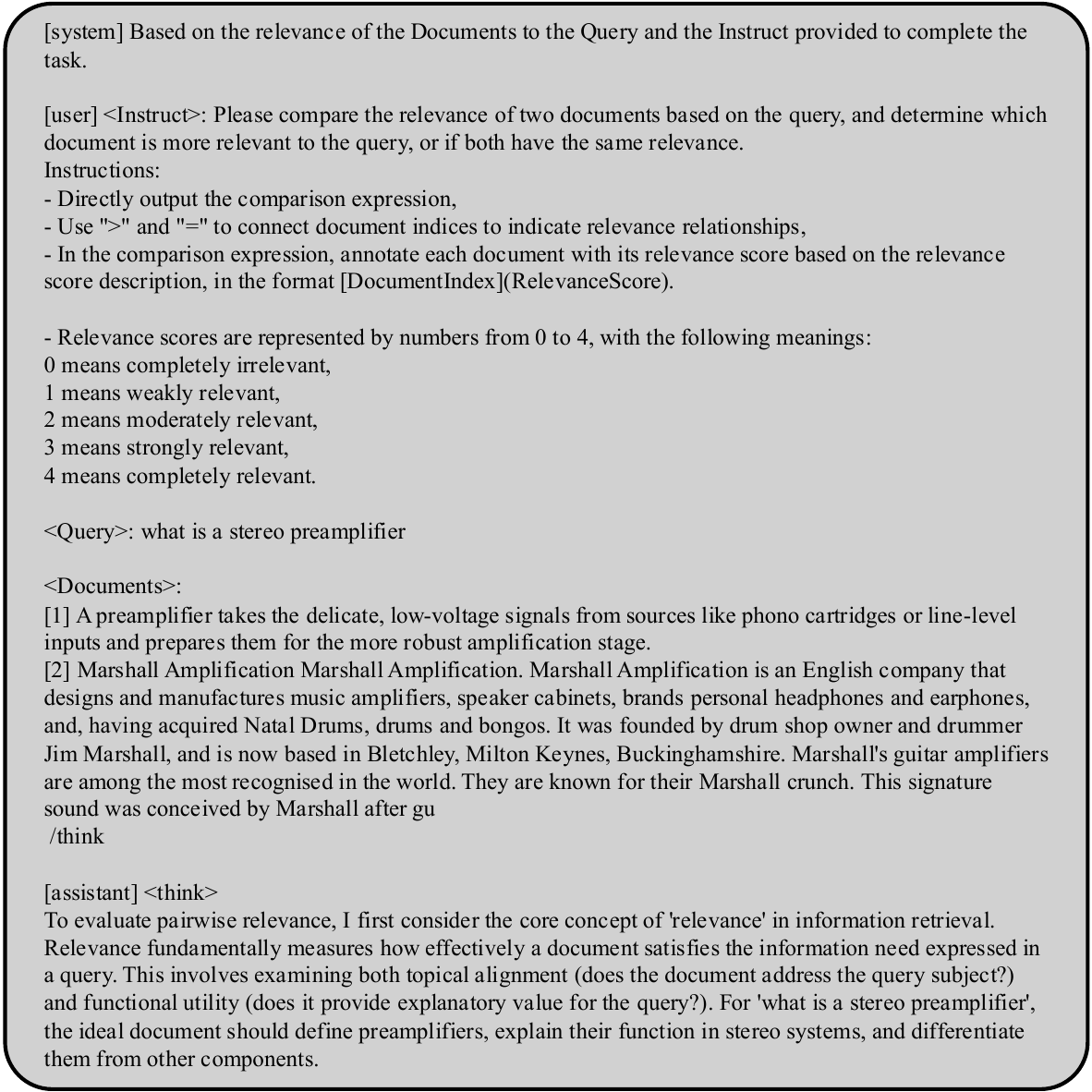} 
\caption{
    Example of a pairwise fine-grained training sample in \texttt{messages} format with explicit reasoning (\texttt{/think} mode) in TFRank (1/2).
}
\label{fig:pairwise_fg_think_1}
\end{figure*}

\begin{figure*}[ht]
\centering
\includegraphics[width=0.75\linewidth]{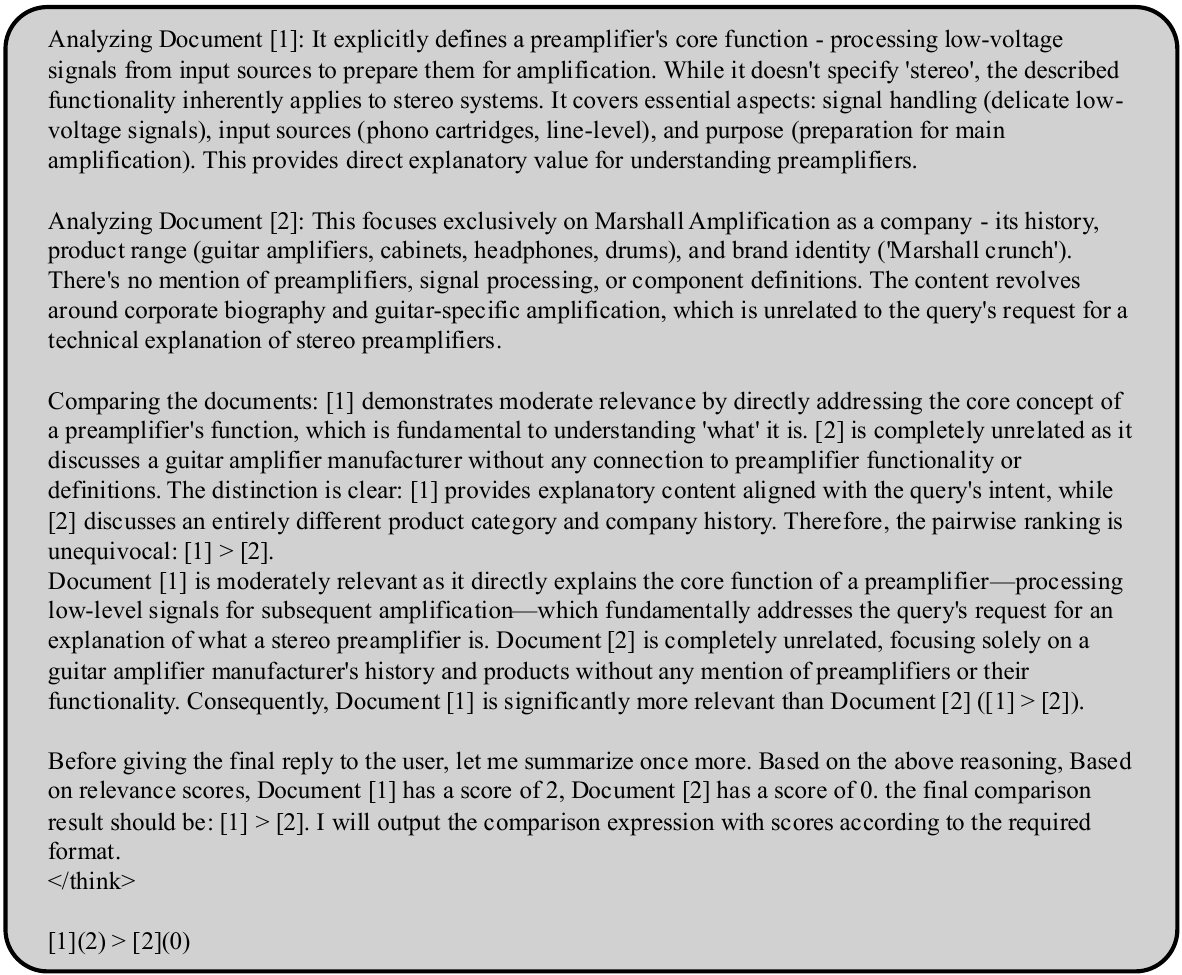} 
\caption{
    Example of a pairwise fine-grained training sample in \texttt{messages} format with explicit reasoning (\texttt{/think} mode) in TFRank (2/2).
}
\label{fig:pairwise_fg_think_2}
\end{figure*}

\begin{figure*}[ht]
\centering
\includegraphics[width=0.75\linewidth]{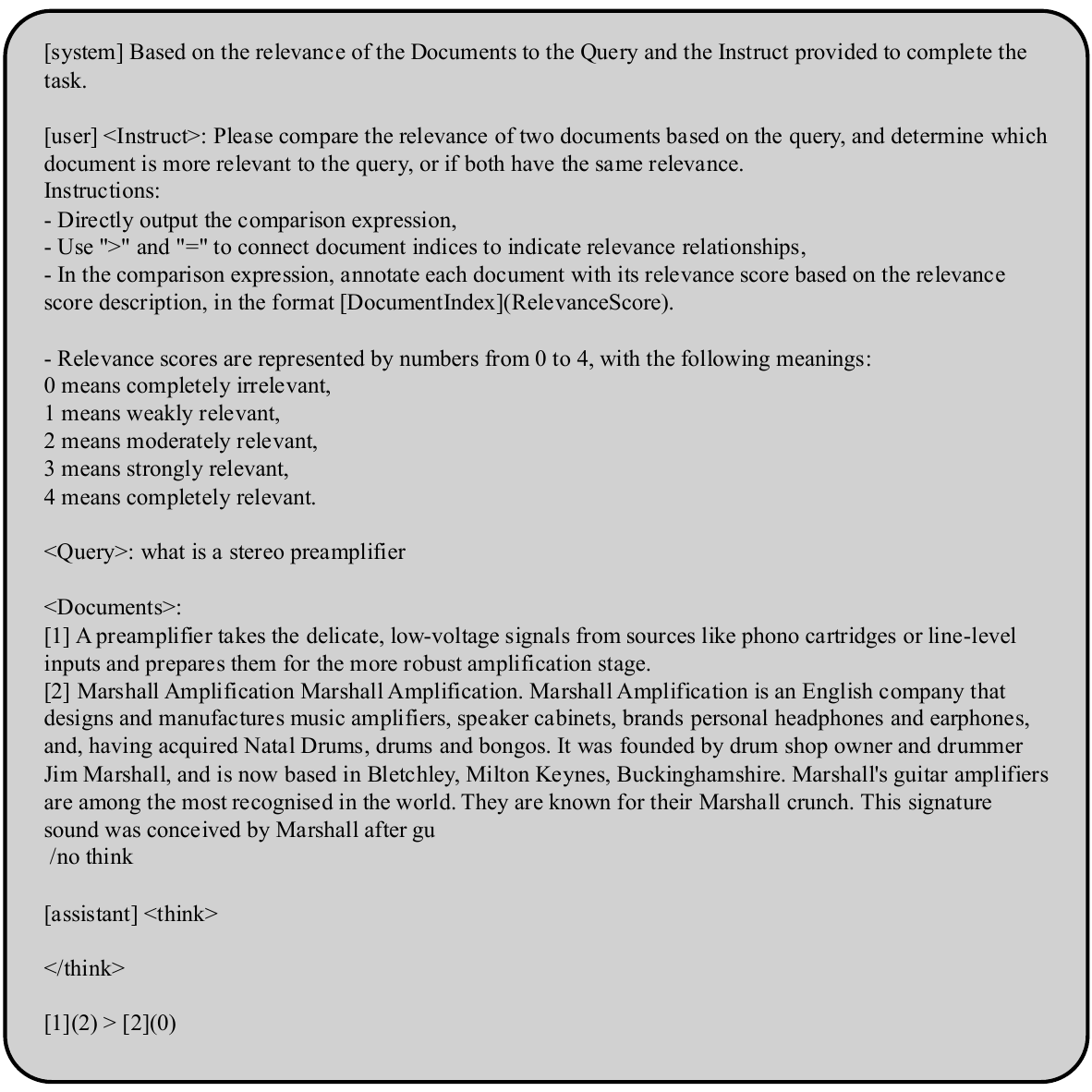} 
\caption{
    Example of a pairwise fine-grained training sample in \texttt{messages} format without explicit reasoning (\texttt{/no think} mode) in TFRank.
}
\label{fig:pairwise_fg_nothink}
\end{figure*}

\begin{figure*}[ht]
\centering
\includegraphics[width=0.75\linewidth]{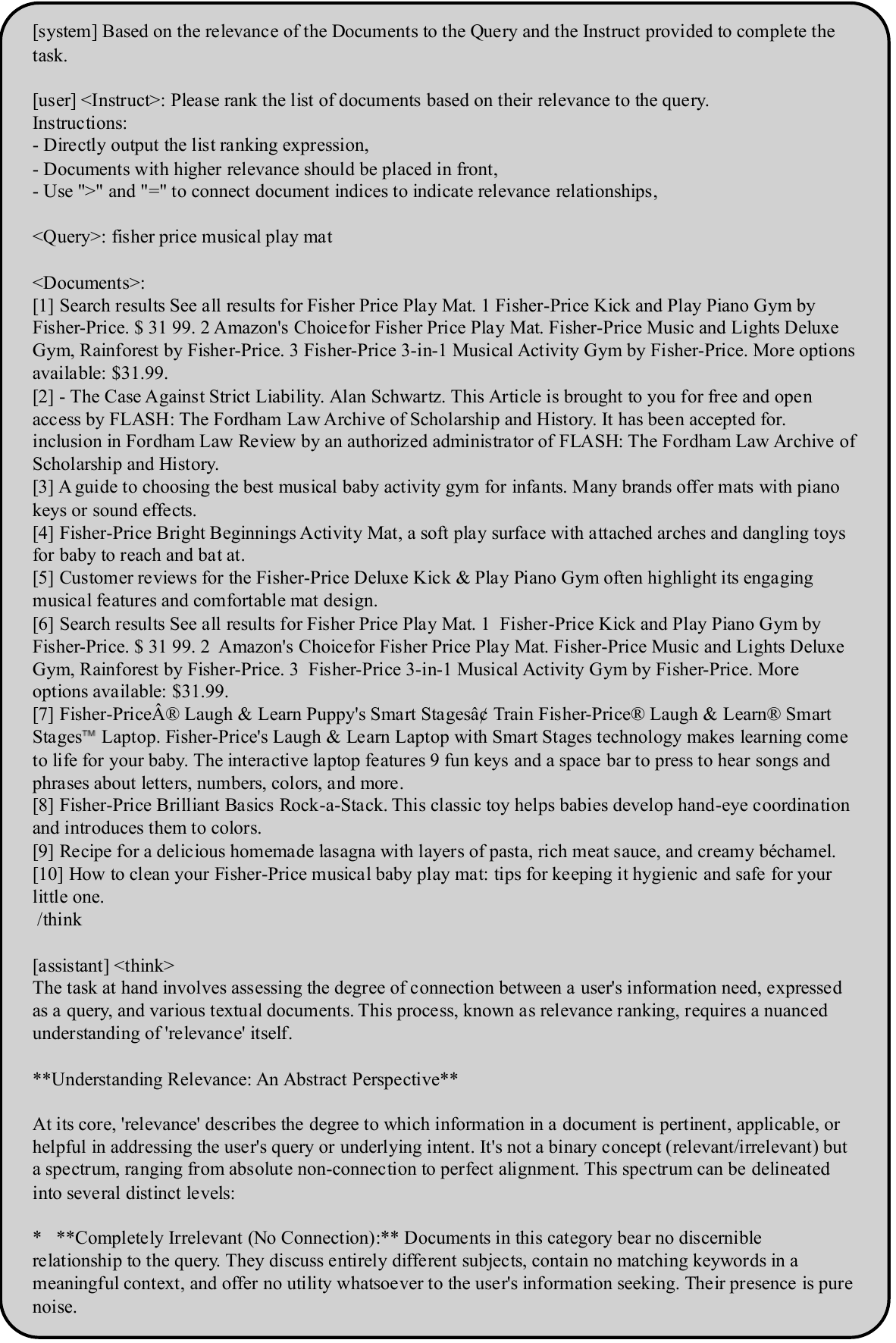} 
\caption{
    Example of a listwise ID-only training sample in \texttt{messages} format with explicit reasoning (\texttt{/think} mode) in TFRank (1/4).
}
\label{fig:listwise_id_think_1}
\end{figure*}

\begin{figure*}[ht]
\centering
\includegraphics[width=0.75\linewidth]{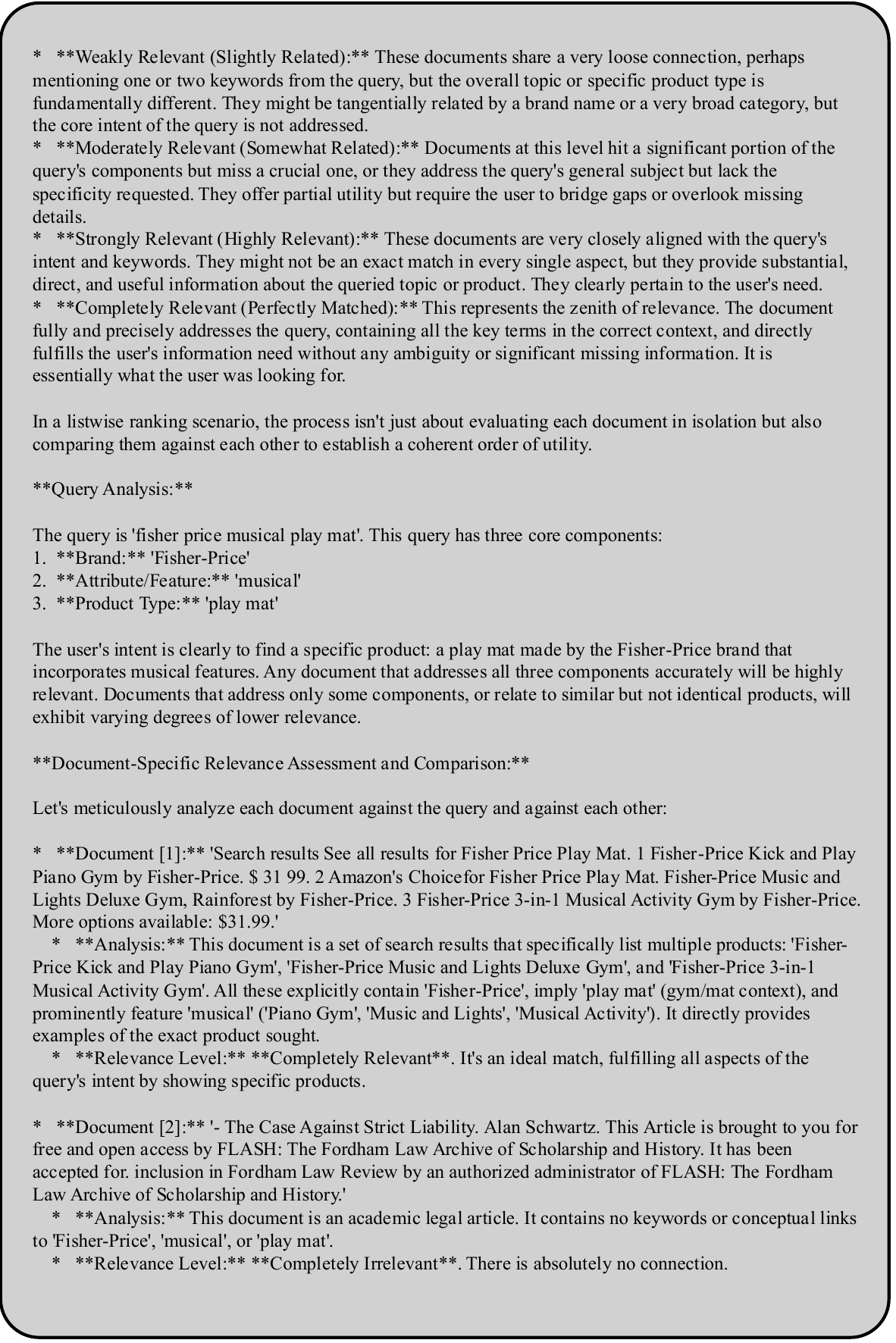} 
\caption{
    Example of a listwise ID-only training sample in \texttt{messages} format with explicit reasoning (\texttt{/think} mode) in TFRank (2/4).
}
\label{fig:listwise_id_think_2}
\end{figure*}

\begin{figure*}[ht]
\centering
\includegraphics[width=0.75\linewidth]{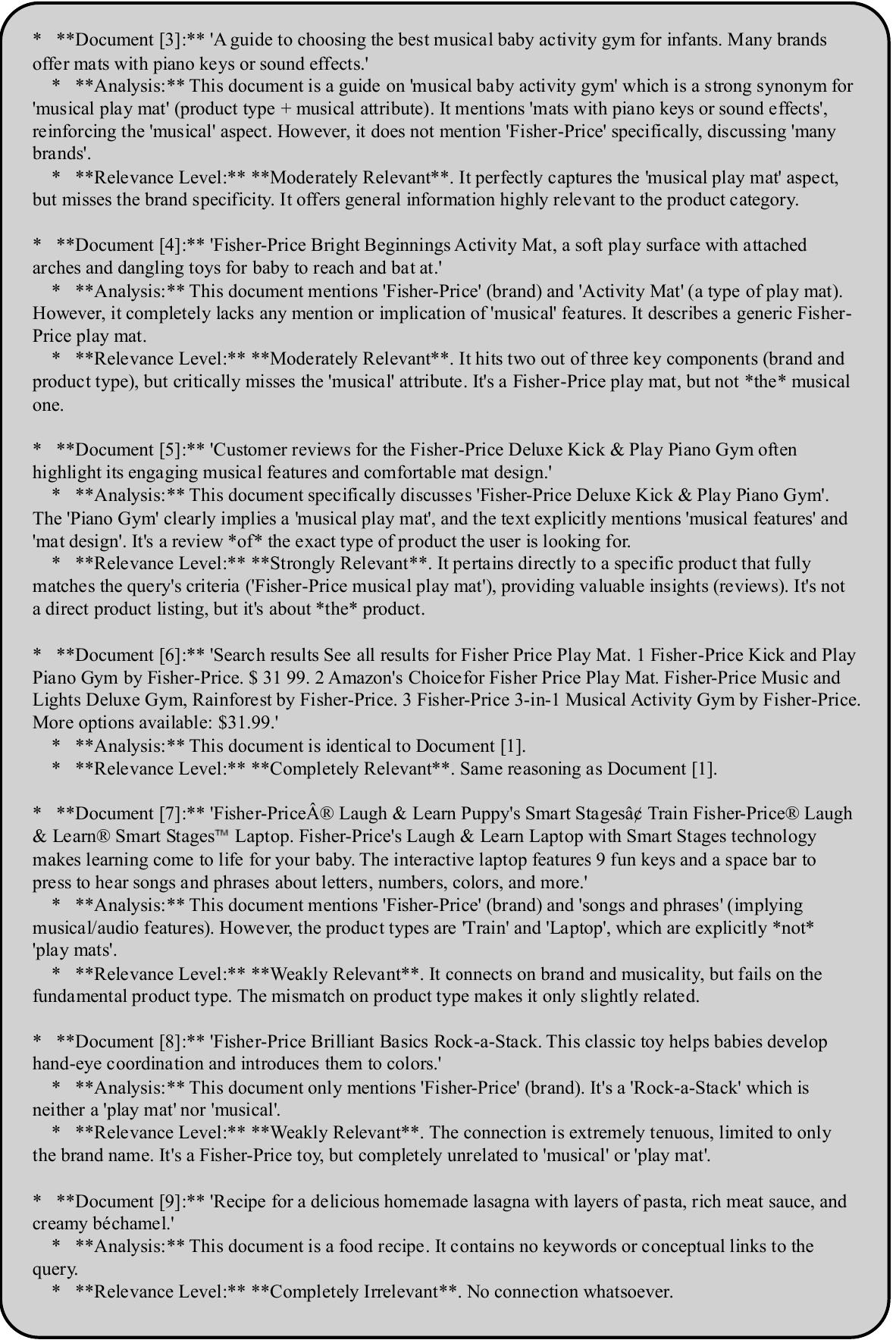} 
\caption{
    Example of a listwise ID-only training sample in \texttt{messages} format with explicit reasoning (\texttt{/think} mode) in TFRank (3/4).
}
\label{fig:listwise_id_think_3}
\end{figure*}

\begin{figure*}[ht]
\centering
\includegraphics[width=0.75\linewidth]{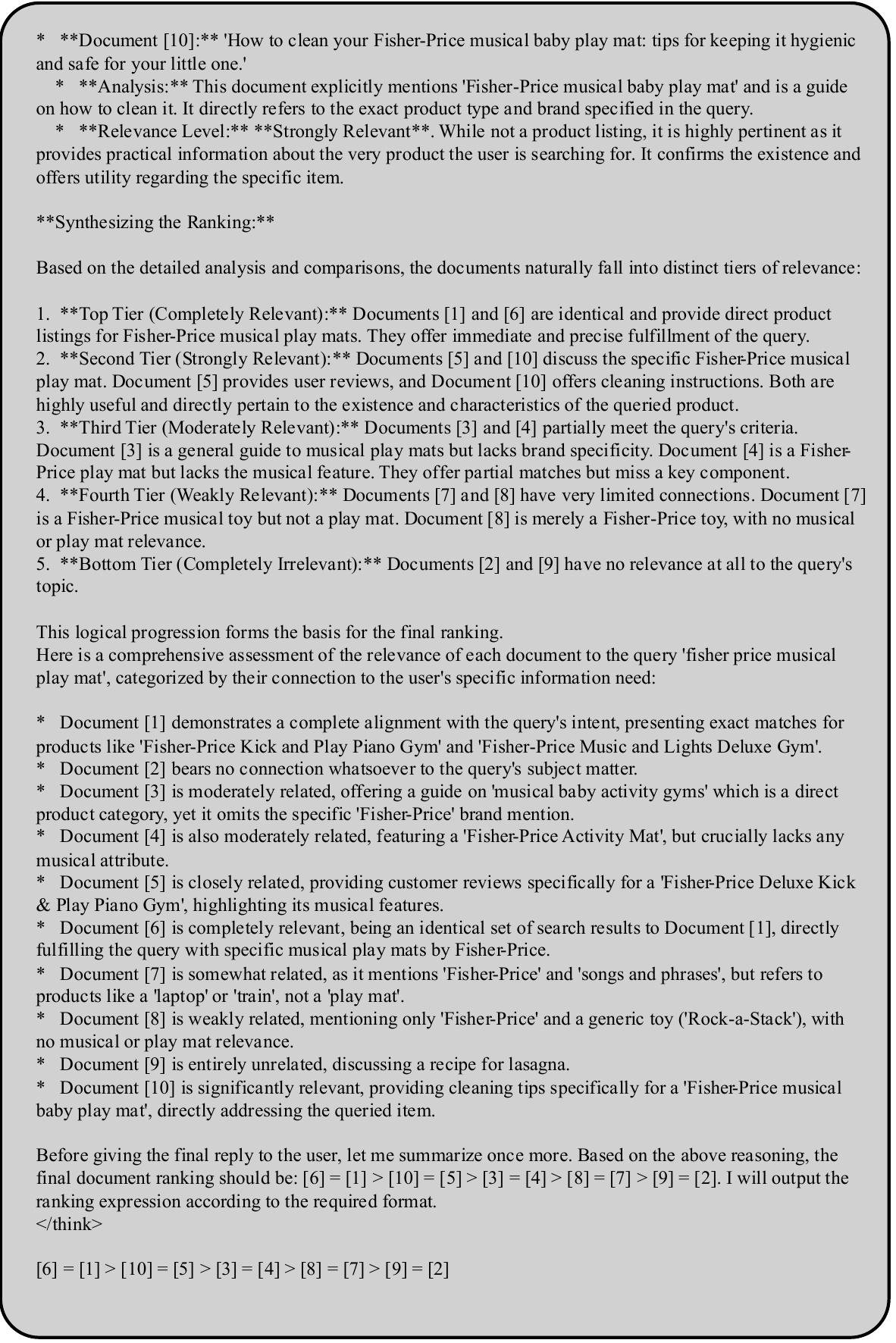} 
\caption{
    Example of a listwise ID-only training sample in \texttt{messages} format with explicit reasoning (\texttt{/think} mode) in TFRank (4/4).
}
\label{fig:listwise_id_think_4}
\end{figure*}

\begin{figure*}[ht]
\centering
\includegraphics[width=0.75\linewidth]{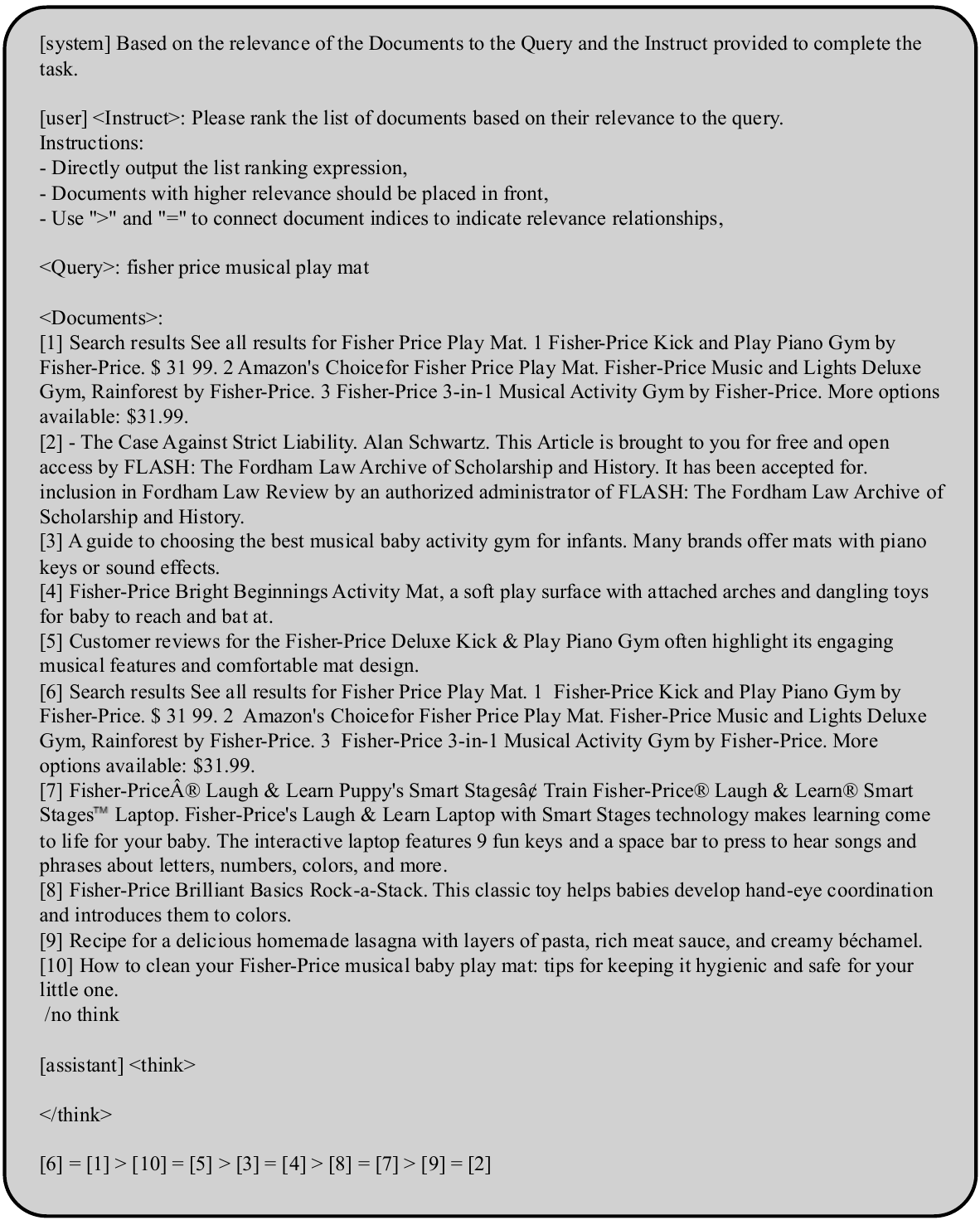} 
\caption{
    Example of a listwise ID-only training sample in \texttt{messages} format without explicit reasoning (\texttt{/no think} mode) in TFRank.
}
\label{fig:listwise_id_nothink}
\end{figure*}

\begin{figure*}[ht]
\centering
\includegraphics[width=0.75\linewidth]{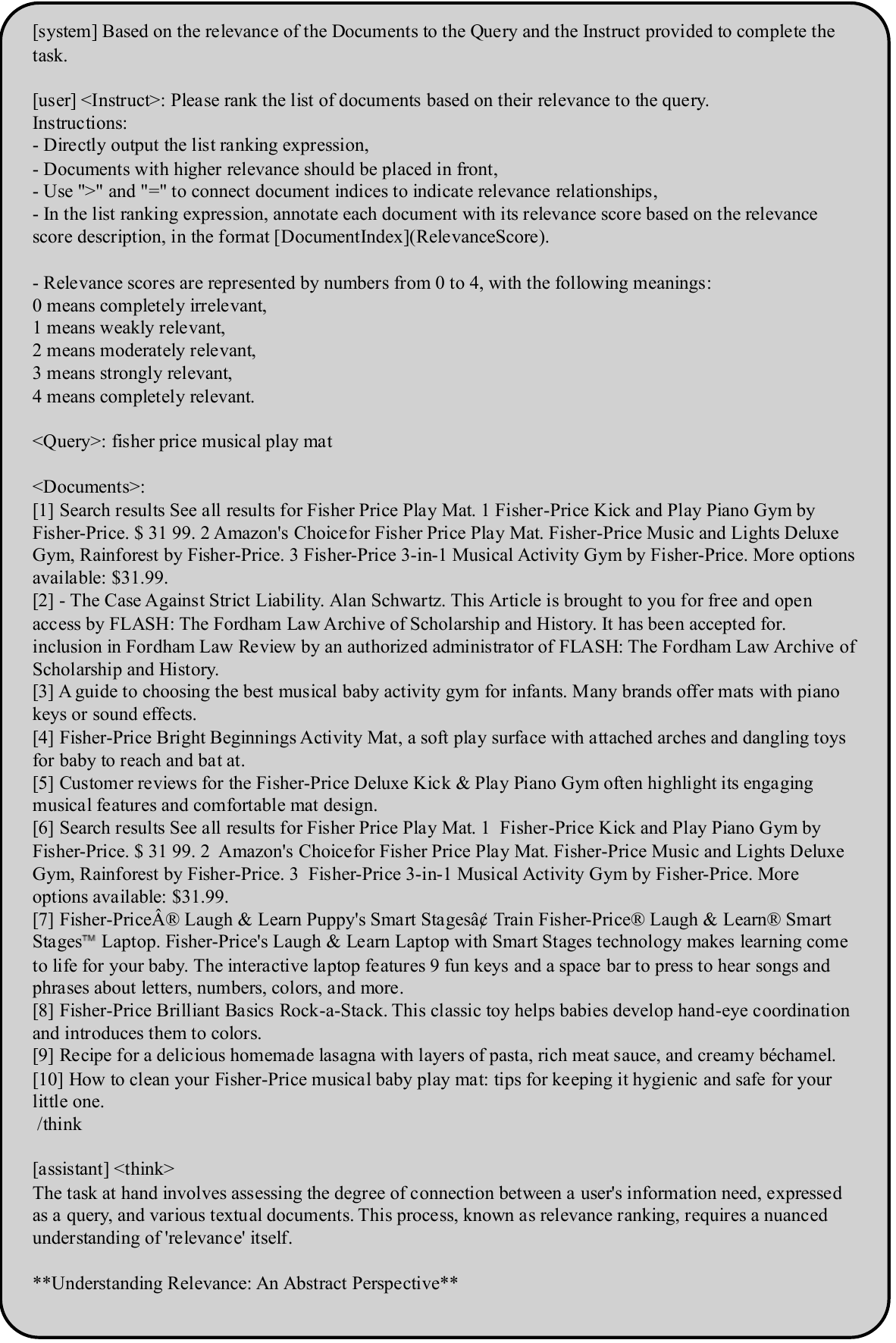} 
\caption{
    Example of a listwise fine-grained training sample in \texttt{messages} format with explicit reasoning (\texttt{/think} mode) in TFRank (1/5).
}
\label{fig:listwise_fg_think_1}
\end{figure*}

\begin{figure*}[ht]
\centering
\includegraphics[width=0.75\linewidth]{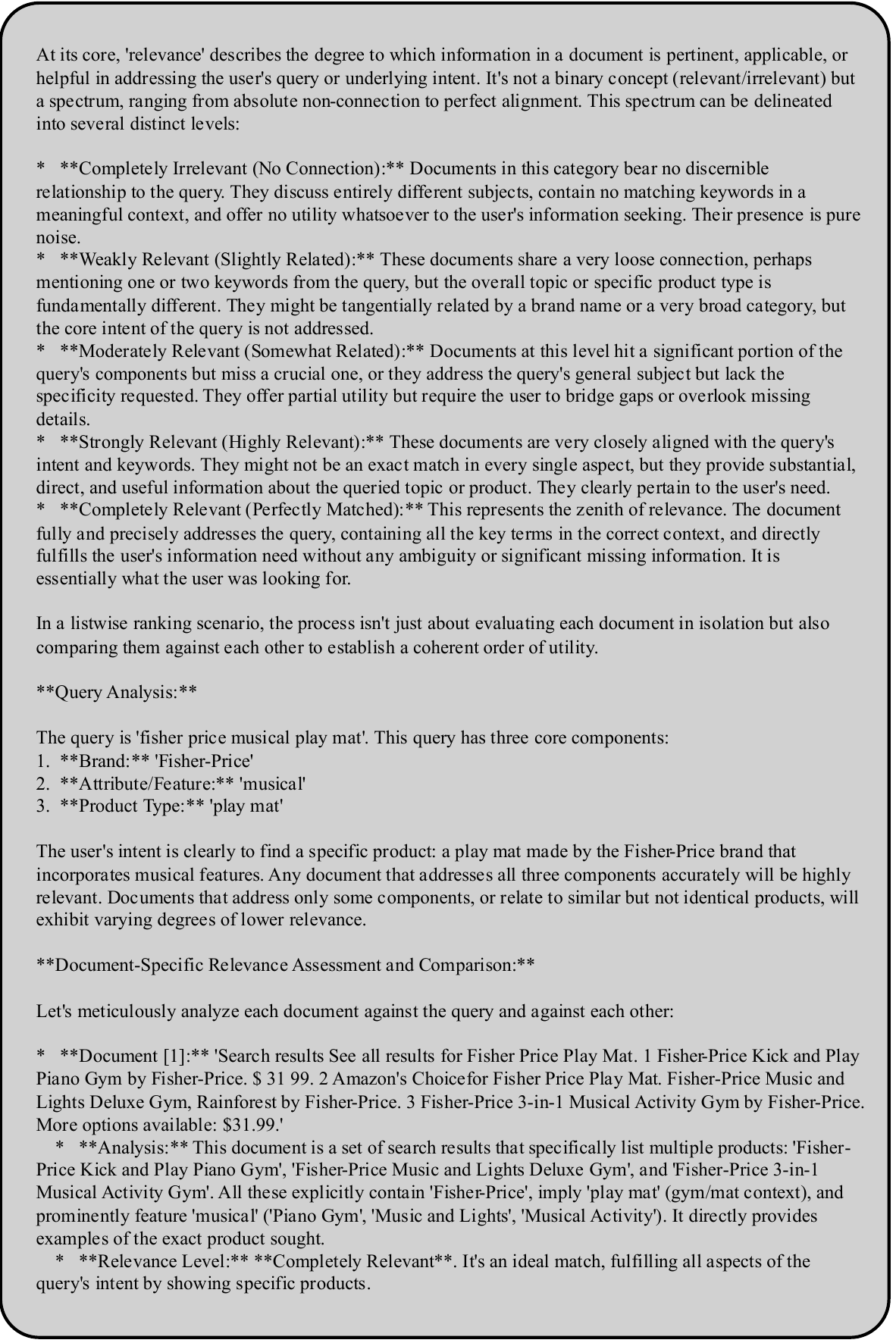} 
\caption{
    Example of a listwise fine-grained training sample in \texttt{messages} format with explicit reasoning (\texttt{/think} mode) in TFRank (2/5).
}
\label{fig:listwise_fg_think_2}
\end{figure*}

\begin{figure*}[ht]
\centering
\includegraphics[width=0.75\linewidth]{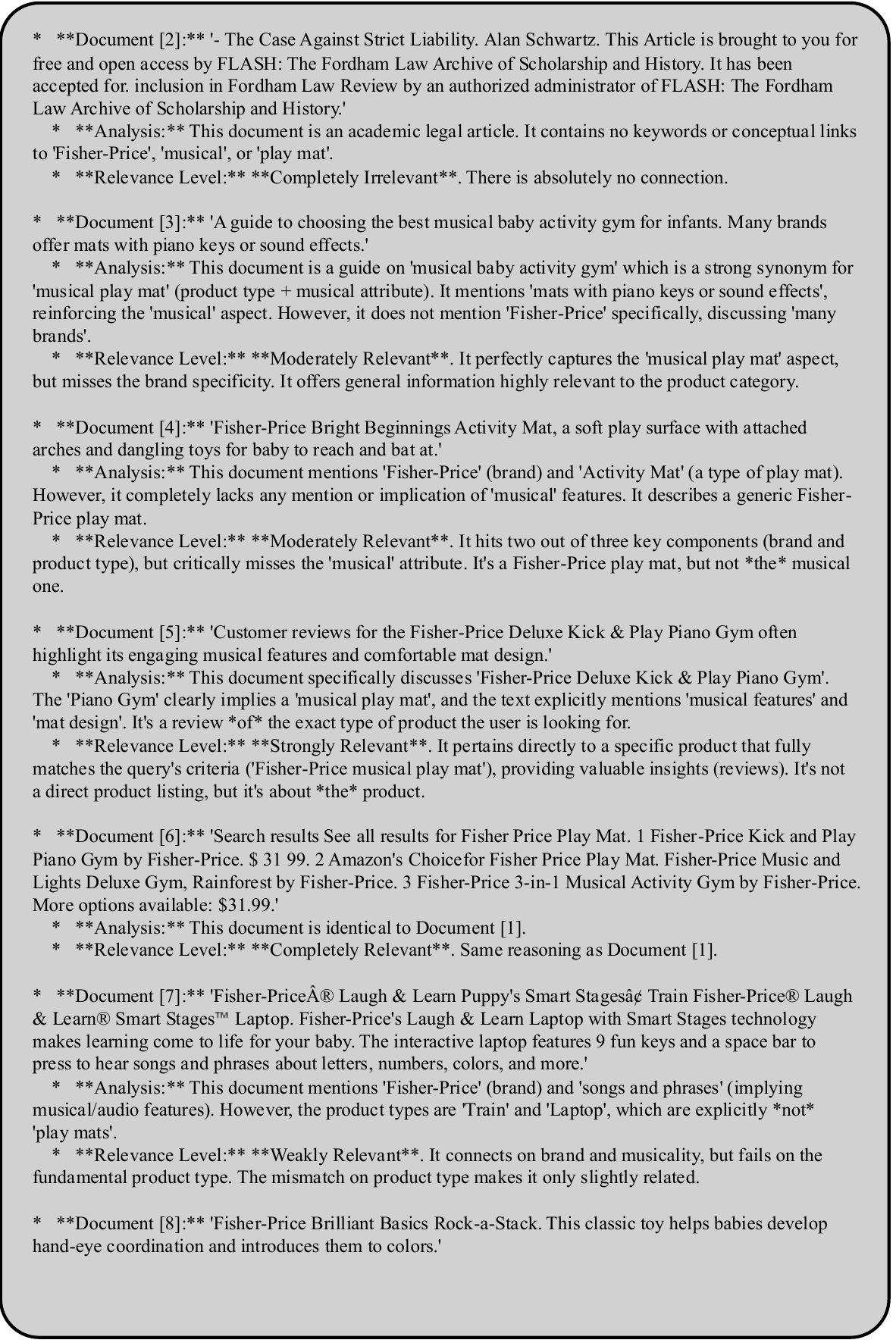} 
\caption{
    Example of a listwise fine-grained training sample in \texttt{messages} format with explicit reasoning (\texttt{/think} mode) in TFRank (3/5).
}
\label{fig:listwise_fg_think_3}
\end{figure*}

\begin{figure*}[ht]
\centering
\includegraphics[width=0.75\linewidth]{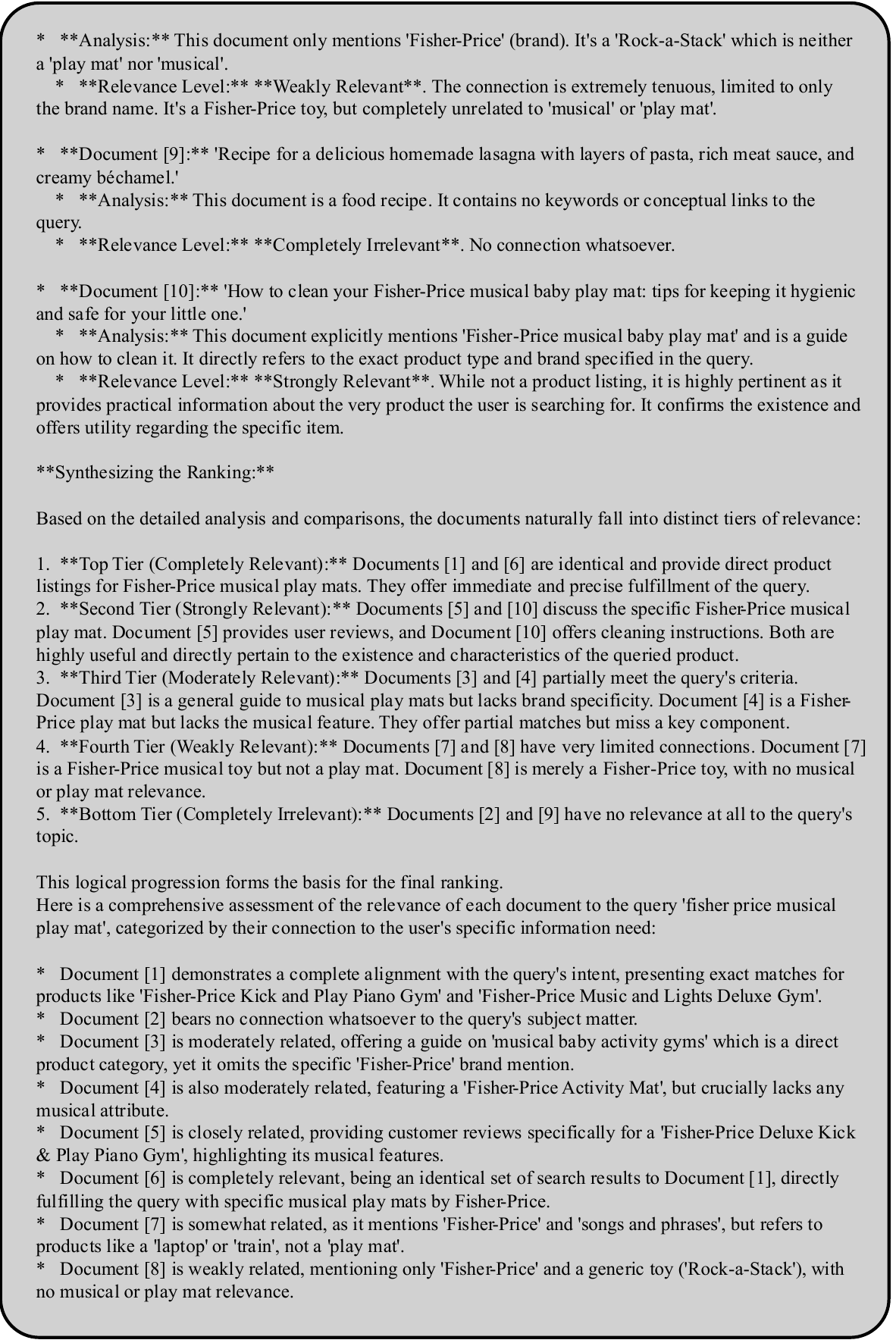} 
\caption{
    Example of a listwise fine-grained training sample in \texttt{messages} format with explicit reasoning (\texttt{/think} mode) in TFRank (4/5).
}
\label{fig:listwise_fg_think_4}
\end{figure*}

\begin{figure*}[ht]
\centering
\includegraphics[width=0.75\linewidth]{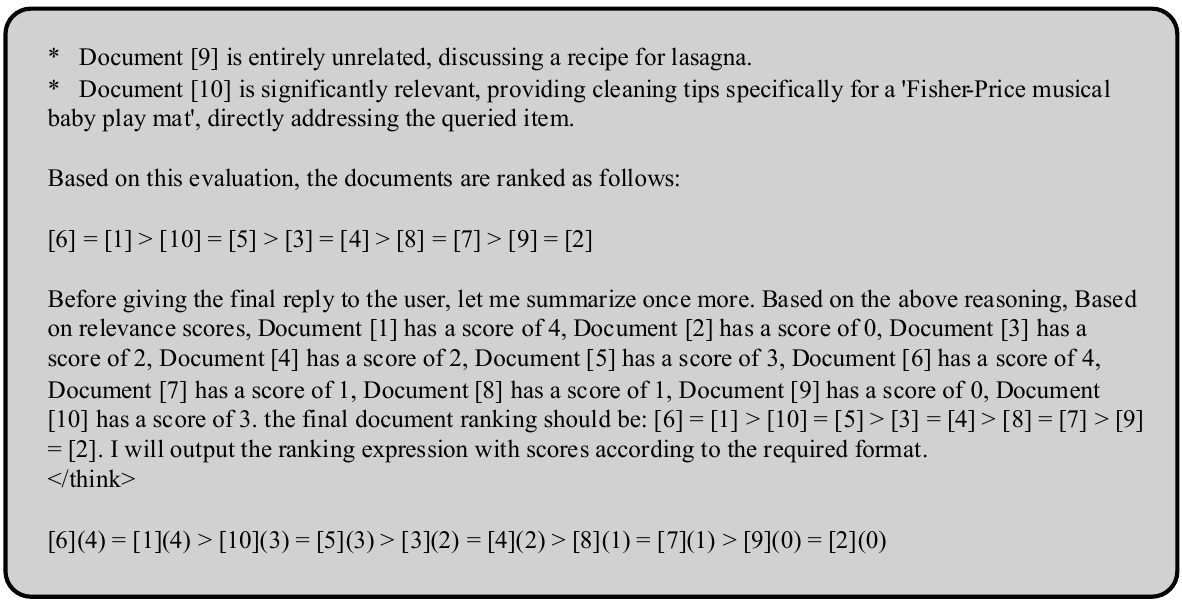} 
\caption{
    Example of a listwise fine-grained training sample in \texttt{messages} format with explicit reasoning (\texttt{/think} mode) in TFRank (5/5).
}
\label{fig:listwise_fg_think_5}
\end{figure*}

\begin{figure*}[ht]
\centering
\includegraphics[width=0.75\linewidth]{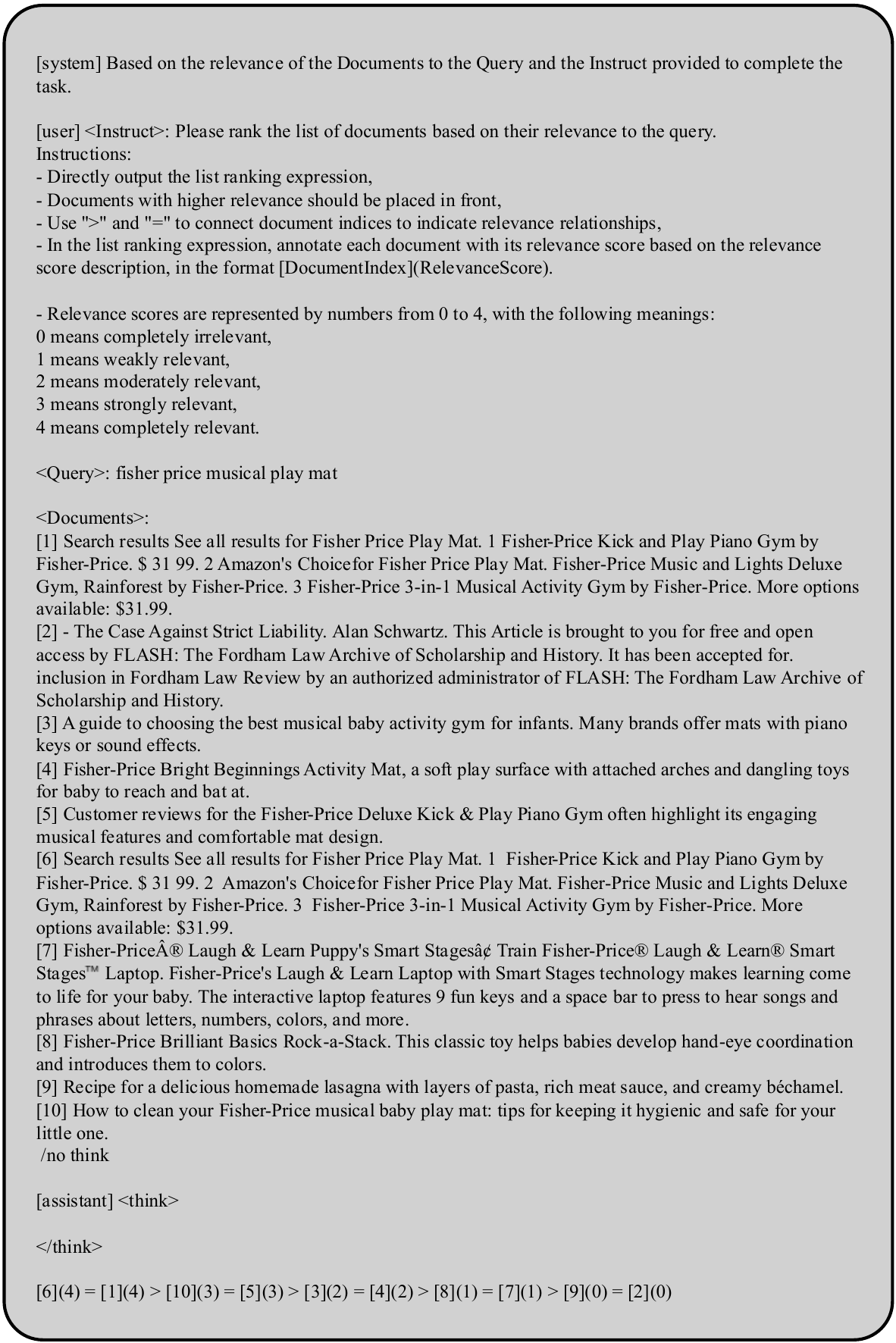} 
\caption{
    Example of a listwise fine-grained training sample in \texttt{messages} format without explicit reasoning (\texttt{/no think} mode) in TFRank.
}
\label{fig:listwise_fg_nothink}
\end{figure*}

From Figure~\ref{fig:inference_template} to Figure~\ref{fig:listwise_fg_nothink}, we provide the exact prompt templates and sample instances for each task paradigm (pointwise, pairwise, listwise), with or without fine-grained label, and different inference modes (\texttt{/think}, \texttt{/no think}).

\renewcommand{\thetable}{B\arabic{table}}
\renewcommand{\thefigure}{B\arabic{figure}}
\setcounter{figure}{0}
\setcounter{table}{0}
\section{Experimental Supplementary Material}
\subsection{Full Experimental Results on BEIR, R2MED, and BRIGHT Leaderboard}
Table~\ref{tab:beir_main_pytrec_eval} presents the complete results(NDCG@10) on the specific BEIR benchmark datasets, which follows the setting of Rank1. Table~\ref{tab:bright_leaderboard_pytrec_eval} presents the complete results(NDCG@10) on the specific BRIGHT Leaderboard. Table~\ref{tab:r2med_main_e5} presents the complete results(NDCG@10) on the specific R2MED benchmark datasets, which follows the setting of ReasonRank.
\subsection{Significance Testing}
To assess the robustness of our findings, we conduct paired significance tests (both paired $t$-test and Wilcoxon signed-rank test) on the per-query NDCG@10 scores for BRIGHT ($n=1384$ queries). Table~\ref{tab:bright_pvalues} reports $p$-values for multiple TFRank variants compared to REARANK-7B, under both BM25 and ReasonIR retrievers. Across all comparisons, both tests consistently yield highly significant $p$-values ($p \ll 0.01$), confirming that the improvements of TFRank over REARANK are statistically robust and not due to random variation.
\subsection{Additional Inference Notes}
For all experiments, we use the vLLM inference framework, a fast and easy-to-use library for LLM inference and serving. During inference, the temperature is set to $0$ to eliminate randomness and ensure deterministic outputs. The specific inference completion prompt template is shown in Figure~\ref{fig:inference_template}.

Due to the current limitations of vLLM, only the top-20 tokens (by logit value) are accessible for extracting model scores (e.g., for yes, no, and the score tokens 0-4). If none of the target tokens appear in the top-20 results, we assign a neutral score of $0.5$ to avoid errors. Notably, for all sufficiently trained TFRank models, such cases are sporadic.
\clearpage

\begin{table*}[ht]
    \centering \small
    \setlength{\tabcolsep}{1.5mm}
    \begin{tabular}{lcc|ccccccccc|c}
        \toprule
        \multicolumn{2}{c}{\textbf{Model}} & \textbf{Approach} & ArguA & ClimF & DBP
        & FiQA & NFCorp & SciDoc 
        & SciFact & Touche & TrecC & \textbf{Avg.} \\
        \midrule
        BM25   & - & Pointwise & \textbf{39.7} & 16.5 & 31.8 & 23.6 & 33.8 & 14.9 & 67.9 & \textbf{44.2} & 59.5 & 36.9 \\
        RankGPT-4 & Zero-shot & Listwise & - & - & \textbf{47.1} & - & \textbf{38.5} & - & 75.0 & 38.6 & \textbf{85.5} & - \\
        Rank1-7B & SFT & Pointwise & 26.4 & 16.2 & 37.7 & 38.4 & 37.9 & 16.5 & 76.1 & 24.5 & 79.5 & 39.2 \\
        Rank1-14B & SFT & Pointwise & 32.2 & 15.6 & 34.2 & 36.6 & 35.1 & 16.6 & 73.8 & 25.9 & 78.0 & 38.7 \\
        REARANK-7B & GRPO & Listwise & 35.6 & 20.6 & 43.5 & 35.8 & 37.9 & 19.2 & 71.9 & \underline{40.2} & 80.1 & 42.8 \\
        Rank-R1-3B & GRPO & Setwise & 31.5 & 22.5 & 41.2 & 29.8 & 31.8 & 16.3 & 64.0 & 27.4 & 75.9 & 37.8 \\
        Rank-R1-7B & GRPO & Setwise & 37.0 & 24.1 & 43.2 & 40.1 & 36.2 & 18.8 & 76.1 & 33.0 & 82.6 & \underline{43.5} \\
        Rank-R1-14B & GRPO & Setwise & 34.4 & 24.2 & \underline{44.0} & \textbf{43.0} & 37.9 & \textbf{19.7} & 77.5 & 29.6 & \underline{83.9} & \textbf{43.8} \\
        \midrule
        \multicolumn{13}{c}{\textbf{Llama3.2 Series}}\\
        \midrule
        TFRank-1B & SFT & Pointwise & 27.4 & 18.5 & 35.4 & 19.5 & 23.6 & 9.7 & 47.3 & 22.6 & 65.4 & 29.9 \\
        TFRank-3B & SFT & Pointwise & 34.2 & 16.6 & 38.1 & 33.3 & 36.2 & 15.8 & 72.3 & 35.1 & 82.5 & 40.5 \\
        TFRank-8B & SFT & Pointwise & \underline{38.5} & 22.7 & 42.6 & 38.0 & 37.5 & 18.3 & 75.6 & 33.0 & 82.5 & 43.2 \\
        \midrule
        \multicolumn{13}{c}{\textbf{Qwen2.5 Series}}\\
        \midrule
        TFRank-0.5B & SFT & Pointwise & 29.7 & 17.8 & 37.0 & 28.3 & 33.6 & 13.2 & 65.1 & 25.6 & 78.7 & 36.6 \\
        TFRank-1.5B & SFT & Pointwise & 36.5 & 17.8 & 38.1 & 31.4 & 35.1 & 15.5 & 71.8 & 32.2 & 77.4 & 39.5 \\
        TFRank-3B & SFT & Pointwise & 33.2 & 18.8 & 38.3 & 33.2 & 36.0 & 15.6 & 71.8 & 32.0 & 78.0 & 39.7 \\
        TFRank-7B & SFT & Pointwise & 35.7 & 20.0 & 38.5 & 34.8 & 37.6 & 19.1 & 75.0 & 33.0 & 80.0 & 41.5 \\
        \midrule
        \multicolumn{13}{c}{\textbf{Qwen3 Series}} \\
        \midrule
        TFRank-0.6B & SFT & Pointwise & 35.4 & 21.8 & 34.3 & 30.5 & 32.8 & 14.3 & 71.8 & 29.8 & 80.7 & 39.0 \\
        TFRank-0.6B & GRPO$^{\ddagger}$ & Pointwise & 13.9 & 23.0 & 35.7 & 25.2 & 34.4 & 13.3 & 68.7 & 21.3 & 74.9 & 34.5 \\
        TFRank-1.7B & SFT & Pointwise & 37.0 & 21.3 & 38.8 & 31.1 & 33.2 & 16.1 & 70.9 & 29.3 & 76.8 & 39.4 \\
        TFRank-1.7B & GRPO$^{\ddagger}$ & Pointwise & 27.6 & 23.5 & 41.2 & 31.5 & 35.9 & 16.9 & 73.0 & 30.7 & 79.8 & 40.0 \\
        TFRank-4B & SFT & Pointwise & 37.2 & 19.7 & 37.9 & 36.2 & \underline{38.3} & 18.3 & 76.6 & 37.6 & 80.5 & 42.5 \\
        TFRank-4B & GRPO$^{\ddagger}$ & Pointwise & 30.7 & \underline{24.9} & 43.9 & 39.2 & 37.6 & \underline{19.3} & \textbf{78.9} & 35.5 & 81.5 & \underline{43.5} \\
        TFRank-8B & SFT & Pointwise & 36.5 & 21.7 & 37.0 & 38.0 & 38.0 & 17.9 & 74.6 & 35.0 & 80.0 & 42.1 \\
        TFRank-8B & GRPO$^{\ddagger}$ & Pointwise & 31.7 & \textbf{25.4} & 43.6 & \underline{40.5} & 37.5 & 18.8 & \underline{77.7} & 31.2 & 83.8 & 43.4 \\
        \bottomrule
    \end{tabular}
    \caption{
        Main evaluation results (NDCG@10) on the BEIR benchmark using BM25 as the retriever. 
        TFRank models are trained on their respective backbones. 
        For each column, the highest value is \textbf{bolded} and the second highest is \underline{underlined}. The meaning of the $^{\ddagger}$ symbol follows that in Table~\ref{tab:bright_yesno_grpo_pytrec_eval}.
    }
    \label{tab:beir_main_pytrec_eval}
\end{table*}

\begin{table*}[t!]
    \centering
    \begin{tabular}{c|c|c}
        \toprule
        Retriever & Ranker & Score \\
        \midrule
        ReasonIR-8B & Rank-R1-32B (Qwen2.5) & 38.8 \\
        ReasonIR-8B & QwenRerank-32B (Qwen2.5) & 36.9 \\
        ReasonIR-8B & TFRank-8B (Qwen3) GRPO$^{\ddagger}$ & 32.0 \\
        ReasonIR-8B & TFRank-7B (Qwen2.5) SFT+GRPO$^{\ddagger}$ & 26.9 \\
        BM25 (pyserini) & Llama-Reasoning-70B (Llama-3.1) & 25.7 \\
        ReasonIR-8B & REARANK-7B (Qwen2.5) & 24.2 \\
        ReasonIR-8B & Rank-R1-7B (Qwen2.5) & 23.2 \\
        BM25 (pyserini) & TFRank-8B (Qwen3) GRPO$^{\ddagger}$ & 23.6 \\
        BM25 (pyserini) & TFRank-7B (Qwen2.5) SFT+GRPO$^{\ddagger}$ & 21.4 \\
        BM25 (pyserini) & GritLM-7B (Mistral) & 19.4 \\
        BM25 (pyserini) & RankGPT4 (gpt-4-0125-preview) & 17.0 \\
        \bottomrule
    \end{tabular}
    \caption{
        BRIGHT leaderboard (NDCG@10), comparing advanced retriever–ranker pipelines. The meaning of the $^{\ddagger}$ symbol follows that in Table~\ref{tab:bright_yesno_grpo_pytrec_eval}
    }
    \label{tab:bright_leaderboard_pytrec_eval}
\end{table*}

\begin{table*}[ht]
    \centering \small
    \setlength{\tabcolsep}{1.8mm}
    \begin{tabular}{lccccccccc}
        \toprule
        \textbf{Models} 
        & \textbf{Biology} & \textbf{Bioin.} & \textbf{MedS.} & \textbf{MedE.} 
        & \textbf{MedD.} & \textbf{PMCT.} & \textbf{PMCC.} & \textbf{IIYiC.} & \textbf{Avg.} \\
        \midrule
        E5-mistral-7B 
        & 18.3 & 41.5 & 41.0 & 6.4 & 11.4 & 19.8 & 31.0 & 21.4 & 23.8 \\
        \midrule
        \multicolumn{10}{c}{\textbf{Non-reasoning reranker}}\\
        \midrule
        RankT5-3B 
        & 13.2 & 32.8 & 23.4 & 2.1 & 4.2 & 0.6 & 14.6 & 12.4 & 12.9 \\
        RankZephyr-7B 
        & 22.9 & 43.1 & 48.2 & 7.0 & 10.5 & 26.6 & 7.8 & 14.6 & 22.6 \\
        \midrule
        \multicolumn{10}{c}{\textbf{Reasoning reranker}}\\
        \midrule
        Rank-R1-7B 
        & 34.0 & 51.6 & 51.0 & \underline{12.8} & \underline{22.0} & 34.8 & 31.7 & 25.1 & 32.9 \\
        Rank-R1-14B 
        & \underline{38.8} & 53.8 & \textbf{57.9} & \textbf{15.1} & \textbf{25.5} & \textbf{40.6} & \textbf{42.5} & \textbf{29.6} & \textbf{38.0} \\
        Rank1-7B 
        & 32.6 & \textbf{55.6} & \underline{54.7} & \underline{12.8} & 20.0 & 34.4 & 30.2 & 18.2 & 32.3 \\
        TFRank-7B SFT+GRPO$^{\ddagger}$ (Qwen2.5) 
        & 29.6 & \textbf{55.6} & 53.5 & 11.4 & 19.3 & \underline{37.1} & 34.2 & 20.3 & 32.6 \\
        TFRank-8B GRPO$^{\ddagger}$ (Qwen3) 
        & \textbf{42.7} & \underline{55.2} & 53.2 & 12.2 & 17.2 & 33.7 & \underline{37.1} & \underline{26.1} & \underline{34.7} \\
        \bottomrule
    \end{tabular}
    \caption{
        Main evaluation results (NDCG@10) on the R2MED benchmark using E5-mistral as the retriever. 
        TFRank models are trained on their respective backbones. 
        For each column, the highest value is \textbf{bolded} and the second highest is \underline{underlined}. The meaning of the $^{\ddagger}$ symbol follows that in Table~\ref{tab:bright_yesno_grpo_pytrec_eval}.
    }
    \label{tab:r2med_main_e5}
\end{table*}

\begin{table*}[ht]
    \centering
    \small
    \begin{tabular}{l l c c}
        \toprule
        \textbf{Retriever} & \textbf{Ranker Comparison} & \textbf{T-test} & \textbf{Wilcoxon} \\
        \midrule
        BM25 (pyserini) & TFRank-7B SFT (Qwen2.5) vs.\ REARANK-7B & 2.75$\times$10$^{-2}$ & 9.07$\times$10$^{-3}$ \\
        BM25 (pyserini) & TFRank-8B SFT (Qwen3) vs.\ REARANK-7B   & 4.95$\times$10$^{-8}$ & 1.69$\times$10$^{-8}$ \\
        BM25 (pyserini) & TFRank-8B GRPO$^{\ddagger}$ (Qwen3) vs.\ REARANK-7B & 1.74$\times$10$^{-17}$ & 1.12$\times$10$^{-17}$ \\
        ReasonIR-8B & TFRank-7B SFT (Qwen2.5) vs.\ REARANK-7B & 1.91$\times$10$^{-5}$ & 3.21$\times$10$^{-5}$ \\
        ReasonIR-8B & TFRank-8B SFT (Qwen3) vs.\ REARANK-7B & 1.06$\times$10$^{-6}$ & 9.22$\times$10$^{-8}$ \\
        ReasonIR-8B & TFRank-8B GRPO$^{\ddagger}$ (Qwen3) vs.\ REARANK-7B & 6.03$\times$10$^{-17}$ & 3.34$\times$10$^{-17}$ \\
        \bottomrule
    \end{tabular}
    \caption{
        Some paired significance tests ($p$-values) for NDCG@10 on BRIGHT ($n=1384$ queries). Each row compares TFRank and REARANK-7B under the specified retriever and ranker configuration. Both paired $t$-test and Wilcoxon signed-rank test are reported. 2The meaning of the $^{\ddagger}$ symbol follows that in Table~\ref{tab:bright_yesno_grpo_pytrec_eval}.
    }
    \label{tab:bright_pvalues}
\end{table*}

\end{document}